\documentclass[sigconf]{acmart}

\usepackage[linesnumbered,ruled,vlined]{algorithm2e}
\usepackage{pifont}
\usepackage{float}
\usepackage{graphicx} 
\usepackage{tabularx}
\usepackage{enumitem}
\usepackage{bbding}
\usepackage{caption}
\usepackage{subcaption} 
\usepackage{amsmath}
\DeclareMathOperator*{\argmax}{arg\,max}
\DeclareMathOperator*{\argmin}{arg\,min}
\newtheorem*{remark}{Remark}

\AtBeginDocument{%
  \providecommand\BibTeX{{%
    \normalfont B\kern-0.5em{\scshape i\kern-0.25em b}\kern-0.8em\TeX}}}

\setcopyright{acmcopyright}
\copyrightyear{2018}
\acmYear{2018}
\acmDOI{XXXXXXX.XXXXXXX}

\acmPrice{15.00}
\acmISBN{978-1-4503-XXXX-X/18/06}

\begin{document}
\sloppy

\title[Practical and Asymptotically Optimal Quantization of High-Dim. Vectors in Euclidean Space for ANN Search]{Practical and Asymptotically Optimal Quantization of High-Dimensional Vectors in Euclidean Space for Approximate Nearest Neighbor Search}

\author{Jianyang Gao}
\affiliation{%
  \institution{Nanyang Technological University}
  \country{Singapore}}
\email{jianyang.gao@ntu.edu.sg}

\author{Yutong Gou}
\affiliation{%
  \institution{Nanyang Technological University}
  \country{Singapore}}
\email{yutong003@e.ntu.edu.sg}

\author{Yuexuan Xu}
\affiliation{%
  \institution{Nanyang Technological University}
  \country{Singapore}}
\email{yuexuan001@e.ntu.edu.sg}

\author{Yongyi Yang}
\affiliation{%
  \institution{University of Michigan}
  \country{USA}}
\email{yongyi@umich.edu}

\author{Cheng Long\textsuperscript{*}}
\affiliation{%
  \institution{Nanyang Technological University}
  \country{Singapore}}
\email{c.long@ntu.edu.sg}

\author{Raymond Chi-Wing Wong}
\affiliation{%
  \institution{The Hong Kong University of Science and Technology}
  \country{Hong Kong}}
\email{raywong@cse.ust.hk}
\renewcommand{\shortauthors}{Gao, et al.}

\thanks{\textsuperscript{*}Corresponding author.}

\begin{abstract}
\emph{Approximate nearest neighbor (ANN) query} in high-dimensional Euclidean space is a key operator in database systems. 
For this query, quantization is a popular family of methods developed for compressing vectors and reducing memory consumption.
Recently, a method called \emph{RaBitQ} achieves the state-of-the-art performance among these methods. It produces better empirical performance in both accuracy and efficiency when using the same compression rate and provides rigorous theoretical guarantees.
However, the method is only designed for compressing vectors at high compression rates (32x) and lacks support for achieving higher accuracy by using more space. In this paper, we introduce a new quantization method to address this limitation by extending RaBitQ.
The new method inherits the theoretical guarantees of RaBitQ and achieves the asymptotic optimality in terms of the trade-off between space and error bounds as to be proven in this study. 
Additionally, we present efficient implementations of the method, enabling its application to ANN queries to reduce both space and time consumption.
Extensive experiments on real-world datasets confirm that our method consistently outperforms the state-of-the-art baselines in both accuracy and efficiency when using the same amount of memory. 
\end{abstract}

\maketitle
\pagestyle{plain}

\makeatletter
\pretocmd{\@chooseSymbol}{\raisebox{-.5ex}[\height][0pt]}{}{}
\makeatother

\section{Introduction}
\label{sec: introduction}

\emph{Nearest neighbor (NN) queries} for vectors in high-dimensional Euclidean space are a fundamental operator in database systems~\cite{milvus,PASE,apple,Qdrant,elasticsearch,pgvector,pgvectorrs,singlestore}, with a wide range of applications such as retrieval-augmented generation~\cite{rag1} and information retrieval~\cite{colbert,colbertv2}. However, due to the curse of dimensionality~\cite{indyk1998approximate, vafile}, performing exact NN queries on large-scale databases becomes impractical because of their long response time. To balance time and accuracy, researchers often turn to a relaxed alternative: the \emph{approximate nearest neighbor (ANN) query}~\cite{datar2004locality, muja2014scalable, jegou2010product, malkov2018efficient, ge2013optimized}.

In order to respond to user queries with low latency and high accuracy, the majority of existing studies focus on the in-memory setting of ANN which assumes that a device has sufficiently large RAM and mainly targets to optimize the search performance (the time-accuracy trade-off)~\cite{malkov2018efficient,annbenchmark,fu2019fast,li2019approximate,scann}. 
However, in real-world systems, main memory is a precious resource. 
For example, in database systems deployed on commodity PCs~\cite{milvuslite}, clouds~\cite{cloud2024,milvus,manu} and mobile devices~\cite{objectbox,couchbase}, memory consumption non-trivially affects the costs of services and the experience of users. 
Therefore, handling ANN queries with the minimal space usage while ensuring strong search performance (i.e., achieving competitive efficiency at 90\%, 95\% and 99\% recall) has become a crucial challenge.
In particular, the space consumption in ANN queries is 
comprised of two parts, one for the vectors and the other for the index.  
The part for vectors usually dominates the overall consumption because in most applications, they are embeddings produced by deep neural networks~\cite{colbert,openaiembedding}. Each usually has hundreds or thousands of dimensions and is correspondingly represented by hundreds or thousands of floating-point numbers. 
Thus, many existing studies target to reduce the memory consumption by compressing the vectors~\cite{blink_scalar_quantization_graph,jegou2010product,ge2013optimized,rabitq,lsq++,additivePQ}. 

To address this need, a well-known line of studies called \emph{product quantization (PQ)} has been proposed~\cite{jegou2010product,additivePQ,ge2013optimized,lsq++,surveyl2hash}.
During indexing, these methods first construct a \emph{quantization codebook}. For each vector in the database, they find the nearest vector in the codebook as its \emph{quantized vector}, which is represented and stored as a short \emph{quantization code}.
During querying, they estimate the distances between data and query vectors based on the quantization codes.
However, as has been widely reported~\cite{ge2013optimized,jegou2010product,lsq++}, these methods can only produce poor recall (e.g., <80\%) due to their severely lossy compression (i.e., they mostly adopt a large compression rate of $\ge$32x).
To recover the recall, PQ and its variants usually adopt a re-ranking strategy~\cite{surveyl2hash,douze2024faisslibrary}, which accesses the \emph{raw} vectors and computes exact distances to find the NN. 
However, this entails to store the raw vectors in main memory~\footnote{Another thread of studies access the raw vectors stored on disks for re-ranking~\cite{diskann, starling}, which saves memory by making a compromise in efficiency. We note that they are orthogonal to vector compression and are out of the scope of the current study.}, undermining their target of saving space. 
On the other hand, when these methods are applied with a moderate compression rate (e.g., 8x or 4x) in order to produce reasonable recall without re-ranking, they often struggle to deliver competitive performance in terms of both accuracy and efficiency~\cite{blink_scalar_quantization_graph,fastscan} (e.g., they fail to outperform the classical scalar quantization, see Section~\ref{subsec: experimental results}).

A recent study proposes a new scheme of quantization called \emph{RaBitQ}, which outperforms PQ and its variants both empirically and theoretically~\cite{rabitq}. 
Empirically, compared with PQ, when using the same length of the quantization codes, the method produces more accurate distance estimation with less time consumption.
Theoretically, unlike PQ and its variants which have no theoretical guarantee, RaBitQ guarantees that its distance estimation is \emph{unbiased} and has an \emph{asymptotically optimal error bound}.
Despite this, RaBitQ has the limitation that it only supports to compress a vector with a \emph{large} compression rate (i.e., it compresses a $D$-dimensional floating-point vector into a $D$-dimensional binary vector, which corresponds to 32x compression).  
In this case, without re-ranking, the method can hardly produce reasonable recall either~\cite{rabitq}.
Considering the promising performance of RaBitQ at a large compression rate, a natural question is how to extend the method to achieve \emph{moderate} compression rates so as to avoid the re-ranking step which entails storing the \emph{raw} vectors.
Ideally, this extension should (1) maintain accurate distance estimation with both \emph{unbiasedness} and \emph{asymptotic optimality} concerning the trade-off between space and error bounds, and (2) enable efficient distance estimation. 

Nevertheless, to achieve both desiderata, the extension requires non-trivial designs. 
Let $D$ be the dimensionality of a vector.
The original RaBitQ uses $D$ bits to quantize a vector, which corresponds to a large compression rate of 32x. 
To achieve moderate compression rates, we use $(B\cdot D)$ bits to quantize a vector, where $B$ is a small integer.
For example, when $B = 4$ and 8, the compression rates are 8x and 4x respectively.
Correspondingly, we need to construct a codebook with $2^{B\cdot D}$ vectors. 
According to the RaBitQ paper~\cite{rabitq}, (1) the unbiasedness of the estimator holds on condition that the codebook is comprised of randomly rotated unit vectors; and (2) the efficient computation can be achieved because the vectors in the codebook can be represented by binary vectors.
When $B=1$ (which reduces to the case of the original RaBitQ), there exists a natural construction of the codebook, which satisfies both requirements - it constructs a codebook by randomly rotating the vertices of a hypercube which are nested on the unit sphere, i.e., the vectors whose coordinates are $\pm 1/\sqrt{D}$ (see Section~\ref{subsec: pre-rabitq} for more details).
However, when $B > 1$ (which is the case that we target in this paper), there is no such natural construction.
To address the issue, we propose to construct the codebook by shifting, normalizing and randomly rotating the vectors whose coordinates are \emph{$B$-bit unsigned integers}.
The rationale is that (1) the normalization and random rotation operations on these vectors ensure that the codebook is composed of randomly rotated unit vectors, which allows our method to inherit the \emph{unbiasedness} and the \emph{error bound} of RaBitQ;
and (2) the quantization code can be represented by a $D$-dimensional vector of $B$-bit unsigned integers (correspondingly the size of the codebook is $2^{B\cdot D}$) and the distance estimation can be supported by their arithmetic operations without exhaustively decompressing the codes.

Furthermore, for the new quantization codebook above, we note that the task of finding the nearest vector of a data vector (when computing the quantization codes) is not as easy as the original RaBitQ. Therefore, we propose a new efficient algorithm for this task during indexing.
We prove that our method achieves the asymptotic optimality in terms of the trade-off between space and error bounds for the estimation of inner product between unit vectors.
Based on the experimental studies on real-world datasets (Section~\ref{subsec: experimental results}), we also verify that with the same number of bits, our method produces consistently better accuracy than the state-of-the-art method for compressing vectors with a moderate compression rate (e.g., 8x or 4x).

In addition, we apply our method to ANN queries in combination with the \emph{\underline{i}n\underline{v}erted-\underline{f}ile index} (\emph{IVF index})~\cite{jegou2010product}. Beyond trivially using it for distance estimation, we find that the efficiency of the method can be further improved. 
In particular, in a quantization code of our method (a vector of $B$-bit unsigned integers), the concatenation of the most significant bits of all the dimensions is exactly equal to the quantization code of the original RaBitQ (a vector of 0/1 values).
Motivated by this, we split our quantization codes and \emph{separately} store their most significant bits and the remaining bits. 
During querying, we first estimate a distance by accessing only the most significant bits (i.e., it produces exactly the estimated distance based of the original RaBitQ).
If the estimated distance is sufficiently accurate to decide that a data vector cannot be the NN, then we drop it.
Otherwise, we access the remaining bits and incrementally estimate a distance with higher accuracy based on the complete bits. 
Because the distance estimation based on the most significant bits can be realized with a rather efficient SIMD-based implementation called \emph{FastScan}~\cite{fastscanavx2} and the accuracy is sufficient for pruning many data vectors, this operation helps significantly improve the efficiency.

We summarize our major contributions as follows.
\begin{enumerate}
\item We propose a new quantization method by extending RaBitQ. The method constructs the codebook via shifting, normalizing and randomly rotating vectors of $B$-bit unsigned integers. Based on the design, it inherits RaBitQ's unbiased estimator for distance estimation. In addition, we prove that our method achieves the asymptotic optimality in terms of the trade-off between space and error bounds of estimating inner product of unit vectors.

\item We apply our method to ANN queries and introduce the efficient implementation of first estimating a distance based on the most significant bits. When the accuracy is insufficient, we access the remaining bits to incrementally estimate a distance with higher accuracy.

\item We conduct extensive experiments on real-world datasets, which show that (1) our method provides more accurate distance estimation and higher recall than all the baselines on all the tested datasets when using the same number of bits. At a compression rate of about 6.4x and 4.5x, it stably produces over 95\% and 99\% recall respectively without accessing raw vectors for re-ranking; (2) the empirical performance of our method is well aligned with the theoretical analysis.
\end{enumerate}

The remainder of the paper is organized as follows.
Section~\ref{sec:preliminaries} introduces the ANN query and preliminary techniques.
Section~\ref{sec:methodology} presents our method.
Section~\ref{sec:ExRaBitQ for ANN} illustrates the application of RaBitQ to the in-memory ANN search.
Section~\ref{sec:experiments} provides extensive experimental studies on real-world datasets.
Section~\ref{sec:related work} discusses related work.
Section~\ref{sec:conclusion and discussion} presents the conclusion and discussion.

\section{Preliminaries}
\label{sec:preliminaries}

\subsection{ANN Query}
Consider that there is a database which stores $N$ data vectors in the $D$-dimensional Euclidean space. 
The \emph{nearest neighbor (NN) query} targets to find the nearest data vector from the database for a given query vector $\mathbf{q}$. 
Due to the curse of dimensionality, the query is often relaxed to the \emph{approximate nearest neighbor (ANN) query}, which targets smaller time/space consumption by making a slight compromise on the accuracy (e.g., it targets to reach 90\%, 95\% or 99\% recall).
In addition, the ANN query is often extended to finding the $K$ nearest data vectors. 
For the ease of narrative, we assume that $K=1$ in the algorithm description, while we note that our methods can be trivially applied to the query with any arbitrary $K$'s.
We focus on the in-memory ANN. 
However, unlike most of the existing studies on in-memory ANN which assume that the raw data vectors are stored in RAM~\cite{annbenchmark,li2019approximate,malkov2018efficient}, our method targets to reduce the memory consumption by compressing the raw data vectors and storing only the \emph{compressed vectors} in main memory. That is, we target the setting where the raw vectors \emph{cannot} be accessed during querying so as to save the main memory consumption, and this setting is the same as the one studied in~\cite{blink_scalar_quantization_graph}.

\subsection{The Quantization Method RaBitQ}
\label{subsec: pre-rabitq}
A recent paper proposes a new quantization method called \emph{RaBitQ}~\cite{rabitq}, which quantizes a $D$-dimensional real vector into a $D$-bit string. It provides an unbiased estimator of squared distances and guarantees that the estimator has an asymptotically optimal error bound, which always holds regardless of the data distribution. 

Specifically, given a raw data vector $\mathbf{o}_r$ and a raw query vector $\mathbf{q}_r$, it first normalizes the vectors based on a vector $\mathbf{c}$ (e.g., the centroid of a set of data vectors). 
Let $ \mathbf{o}:= \frac{\mathbf{o}_r- \mathbf{c} }{\| \mathbf{o}_r-\mathbf{c} \|}$ and $\mathbf{q}:= \frac{\mathbf{q}_r-\mathbf{c} }{\| \mathbf{q}_r-\mathbf{c} \|}$ be the \emph{normalized} data and query vectors. 
The (squared) Euclidean distance between $\mathbf{o}_r$ and $\mathbf{q}_r$ can be expressed as follows.
\begin{align}
    &\| \mathbf{o}_r-\mathbf{q}_r \|^2 
    = \| {(\mathbf{o}_r - \mathbf{c} )-( \mathbf{q}_r -\mathbf{c} )} \| ^2 
    \\= &\| {\mathbf{o}_r -\mathbf{c} }\|^2 + \| \mathbf{q}_r- \mathbf{c} \|^2 -2 \cdot \| \mathbf{o}_r-\mathbf{c} \| \cdot \| \mathbf{q}_r-\mathbf{c} \|  \cdot \left< \mathbf{q}, \mathbf{o} \right>   \label{eq:dis to ip}
\end{align}
Note that the distance $\| \mathbf{o_r}-\mathbf{c}\|$ can be pre-computed in the index phase and $\| \mathbf{q}_r - \mathbf{c}\|$ can be computed when a query comes and can be shared by many data vectors.
Therefore, the computation of the distances between the raw vectors can be reduced to that of the inner product of their normalized vectors.
Then, it focuses on estimating the inner product of the normalized vectors~\footnote{Without further specification, in this paper, by data and query vectors, we refer to their normalized vectors.}.

During the \underline{\textbf{index}} phase, RaBitQ constructs a set $\mathcal{C}$ of all possible bi-valued unit vectors, each consisting of values of $+ \frac{1}{\sqrt {D} }$ and $- \frac{1}{\sqrt {D}}$. Then, it randomly rotates all vectors in $\mathcal{C}$ by multiplying them with a random \emph{orthogonal matrix}~\cite{johnson1984extensions} (a type of Johnson-Lindenstrauss Transformation) to form a quantization codebook $\mathcal{C}_{r}$.
The process can be described with equations as follows.
\begin{align}
\label{eq: codebook rabitq}
    \mathcal{C}_{r} := \left\{ P \mathbf{x} \ \big| \  \mathbf{x}  \in \mathcal{C}  \right\} , where \ \mathcal{C}:= \left\{ + \frac{1}{\sqrt {D} }, - \frac{1}{\sqrt {D} }  \right\}^D
\end{align}
where $P$ is a random orthogonal matrix~\cite{johnson1984extensions}.
Note that the codebook is solely determined by the random orthogonal matrix $P$ since the set of bi-valued unit vectors is \emph{pre-defined} and does not rely on the data.
Thus, it maintains the codebook $\mathcal{C}_r$ conceptually only
by sampling and storing the matrix $P$. 
For each data vector $\mathbf{o}$, it finds the nearest vector $\mathbf{\bar o}_0$ in $\mathcal{C}_r$ as its \emph{quantized vector}. 
The quantized vector is represented and stored as a \emph{quantization code} $\mathbf{\bar x}_b \in \{0,1\}^D$ (a $D$-bit string) - recall that each quantized vector has a corresponding bi-valued unit vector, denoted by $\mathbf{\bar x}_0$, in $\mathcal{C}$.
Specifically, we have $\mathbf{\bar o}_0=  P \mathbf{\bar x}_0 = P \left( \frac{2}{\sqrt D}\mathbf{\bar x}_b - \frac{1}{\sqrt D} \mathbf{1}_D\right) $ where $\mathbf{1}_D$ is the $D$-dimensional vector whose coordinates are all ones.

During the \underline{\textbf{query}} phase, it constructs an unbiased estimator for the inner product. The estimator has a theoretical error bound. We restate the estimator and its bound as follows.
\begin{lemma}[Restating Theorem 3.2 in \cite{rabitq}]
    \label{lemma: restate}
    $\frac{\left< \mathbf{\bar o}_0, \mathbf{q} \right>}{\left< \mathbf{\bar o}_0, \mathbf{o} \right>}$ is an unbiased estimator of $\left< \mathbf{o,q}\right>$. With the probability of at least $1 - \exp(-c_0\epsilon_0^2)$, its error bound is presented as 
\begin{align}
     \left| \frac{\left< \mathbf{\bar o}_0, \mathbf{q}  \right> }{\left< \mathbf{\bar o}_0,\mathbf{o}  \right> } -\left< \mathbf{o,q} \right>   \right| \le   \sqrt{\frac{{1 - \left< \mathbf{\bar o}_0, \mathbf{o}  \right>^2}}{\left< \mathbf{\bar o}_0, \mathbf{o}  \right>^2 }}\cdot \frac{\epsilon_0}{\sqrt {D-1} } 
\end{align}  
where $c_0$ is a constant and $\epsilon_0$ is a parameter which controls the failure probability of the bound.
\end{lemma}
It is proven that using RaBitQ to quantize a $D$-dimensional vector to a $D$-bit string, the inner product $\left< \mathbf{\bar o}_0, \mathbf{o}\right>$ is highly concentrated around $0.8$~\cite{rabitq}. 
Thus, the above lemma indicates that for estimating the inner product of two $D$-dimensional unit vectors, it guarantees a probabilistic error bound of $O(1/\sqrt D)$ with high probability, which achieves the asymptotic optimality~\cite{2017_focs_additive_error}. 
In terms of the computation of the estimator, we note that $\left< \mathbf{o}, \mathbf{\bar o}_0\right>$ is independent of the query and can be pre-computed before querying. The computation of $\left< \mathbf{q}, \mathbf{\bar o}_0\right>$ can be conducted as follows. 
\begin{align}
    \left< \mathbf{q}, \mathbf{\bar o}_0\right>&=\left< \mathbf{q}, P \left( \frac{2}{\sqrt D}\mathbf{\bar x}_b - \frac{1}{\sqrt D} \mathbf{1}_D\right)\right> \label{eq: sec 2 definition}
    \\ &= \frac{2}{\sqrt{D}} \left< \mathbf{q}',\mathbf{\bar x}_b\right> - \frac{1}{\sqrt{D}} \sum_{i=1}^{D} \mathbf{q}'[i] \label{eq: sec 2 qprime}
\end{align}
Here, $\mathbf{q}'$ denotes $P^{-1}\mathbf{q}$ and $\mathbf{q}'[i]$ denotes the $i$-th dimension of the vector $\mathbf{q}'$;  
(\ref{eq: sec 2 definition}) plugs in the definition of $\mathbf{\bar o}$ and (\ref{eq: sec 2 qprime}) applies $P^{-1}$ on both sides of the inner product.
Note that $\sum_{i=1}^{D} \mathbf{q}'[i]$ depends only on the query vector. Thus, its computation can be conducted once and shared by many data vectors. 
For the computation of $\left< \mathbf{q}',\mathbf{\bar x}_b\right>$, \cite{rabitq} introduces two versions of implementation. One is based on a SIMD-based implementation called \emph{FastScan}~\cite{fastscanavx2}, which can efficiently compute the estimated distances for data vectors batch by batch.
The other is based on bitwise operations, which supports to efficiently estimate distances for individual vectors. 
We refer readers to the original papers of RaBitQ~\cite{rabitq} and FastScan~\cite{fastscanavx2,fastscanpami,fastscan} for more technical details and theoretical analysis.
With all the proposed techniques, RaBitQ supports to unbiasedly estimate the inner product (and further unbiasedly estimate the squared distances) with both promising accuracy and efficiency.

\section{Extending RaBitQ}
\label{sec:methodology}

\subsection{Motivations and Overview}
\label{subsec: challenges}
For ANN query, to avoid storing raw vectors in RAM and reach promising recall at the same time, it is necessary to compress the vectors with a moderate compression rate.
Although RaBitQ achieves an asymptotically optimal error bound when quantizing $D$-dimensional vectors into $D$-bit codes (which corresponds to a rather high compression rate), it is still unclear how it can use more bits in pursuit of higher accuracy. 
In \cite{rabitq}, it presents a simple way to use more bits by padding the vectors. Specifically, it pads $D$-dimensional vectors to $(B\cdot D)$ dimensions with zeros, and thus, it can use $(B\cdot D)$ bits by directly applying RaBitQ. 
Based on Lemma~\ref{lemma: restate}, in this case, it guarantees an error bound of $O(1/(\sqrt {B} \cdot \sqrt{D}))$ with high probability.
However, this result shows that the error decays very slowly with respect to the number of bits used, indicating that this simple extension is not very effective for the scenarios which require a moderate compression rate. 

To be more formal, we note that, with this simple extension, to guarantee an error bound $\epsilon$ with the failure probability of at most $\delta$, it requires $\Theta\left( \frac{1}{\epsilon^2} \log \frac{1}{\delta}\right)$ bits~\footnote{This can be derived from the result in Lemma~\ref{lemma: restate} by plugging in the parameters of $\epsilon$ and $\delta$.}.
However, as has been proven by a theoretical study~\cite{2017_focs_additive_error}, to guarantee an error bound $\epsilon$, it is sufficient (and also necessary) to use $\Theta \left( D\log\left( \frac{1}{D}\cdot \frac{1}{\epsilon^2}\log \frac{1}{\delta}\right)\right)$ bits when the target accuracy is high~\footnote{The result is restated directly from Theorem 4.1 in \cite{2017_focs_additive_error}.}, i.e., when $\frac{1}{\epsilon^2} \log \frac{1}{\delta} > D$.
Note that the required number of bits in this result of the theoretical study is logarithmic to $\epsilon^{-2}$ while the one in the former result of the simple way of padding vectors is linear to $\epsilon^{-2}$.
For instance, when $\frac{1}{\epsilon^2} \log \frac{1}{\delta} = B\cdot D$, the gap would be that between $\Theta \left( D\log B\right)$ and $ \Theta (B\cdot D)$.
This implies that when targeting higher accuracy with more bits used, there is still substantial room to improve from the simple extension of RaBitQ.

Motivated by the discussions above, in this paper, we propose a new quantization method, which extends RaBitQ to support moderate compression rates and achieves asymptotically optimal estimation and efficient computation at the same time. Next, we present the details of our method, including (1) how it quantizes data vectors (Section~\ref{subsec: quantizing data vectors}), (2) how it computes distance estimators (Section~\ref{subsec: computing the estimator}), and (3) its summary and theoretical results (Section~\ref{subsec: summary}).

\subsection{Quantizing Data Vectors}
\label{subsec: quantizing data vectors}

\begin{figure}[thb]
    \vspace{-4mm}
    \centering
    \begin{subfigure}[b]{0.48\linewidth}
        \centering
        \includegraphics[width=\textwidth]{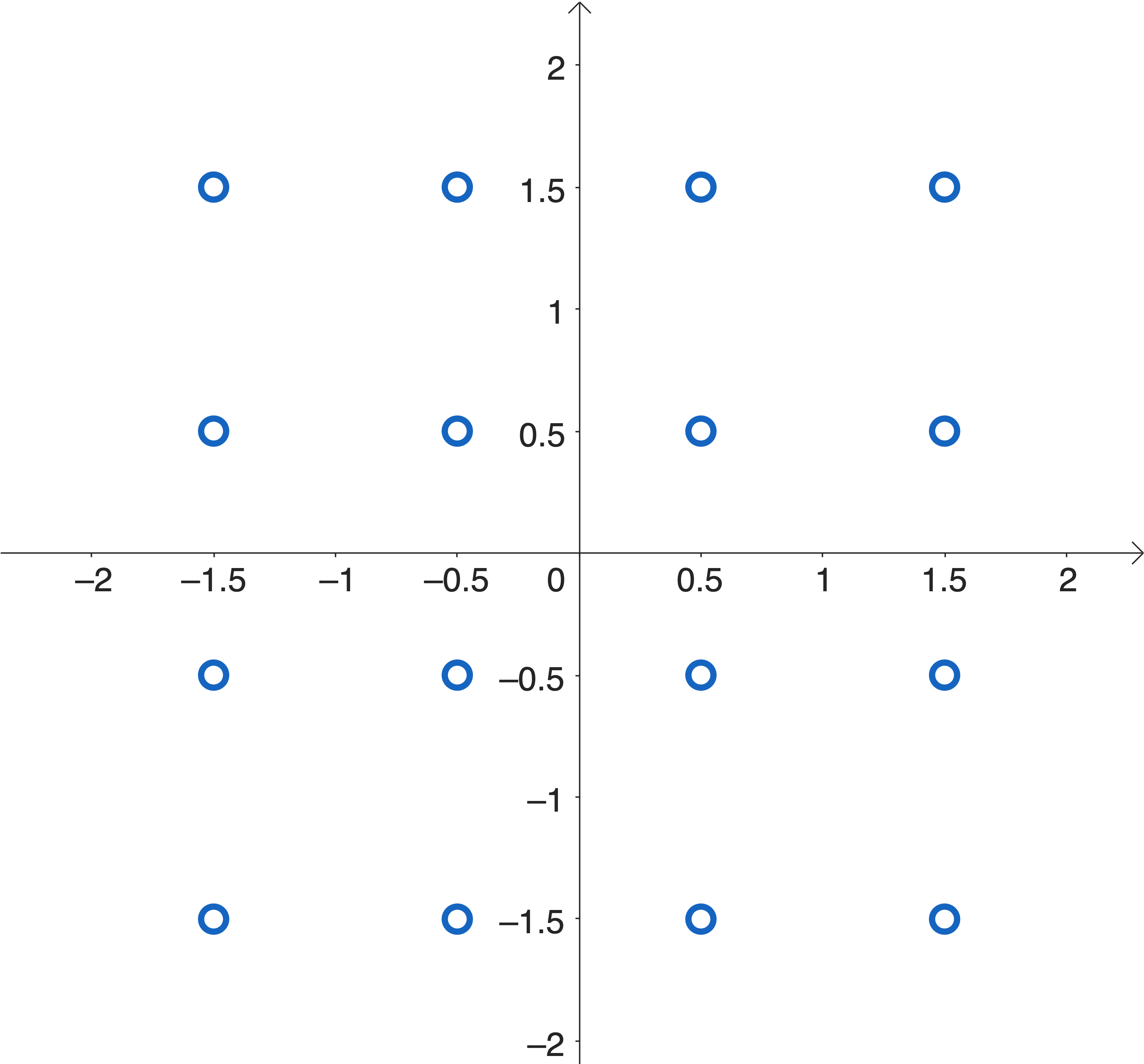}
    \end{subfigure} 
    \begin{subfigure}[b]{0.48\linewidth}
        \centering
        \includegraphics[width=\textwidth]{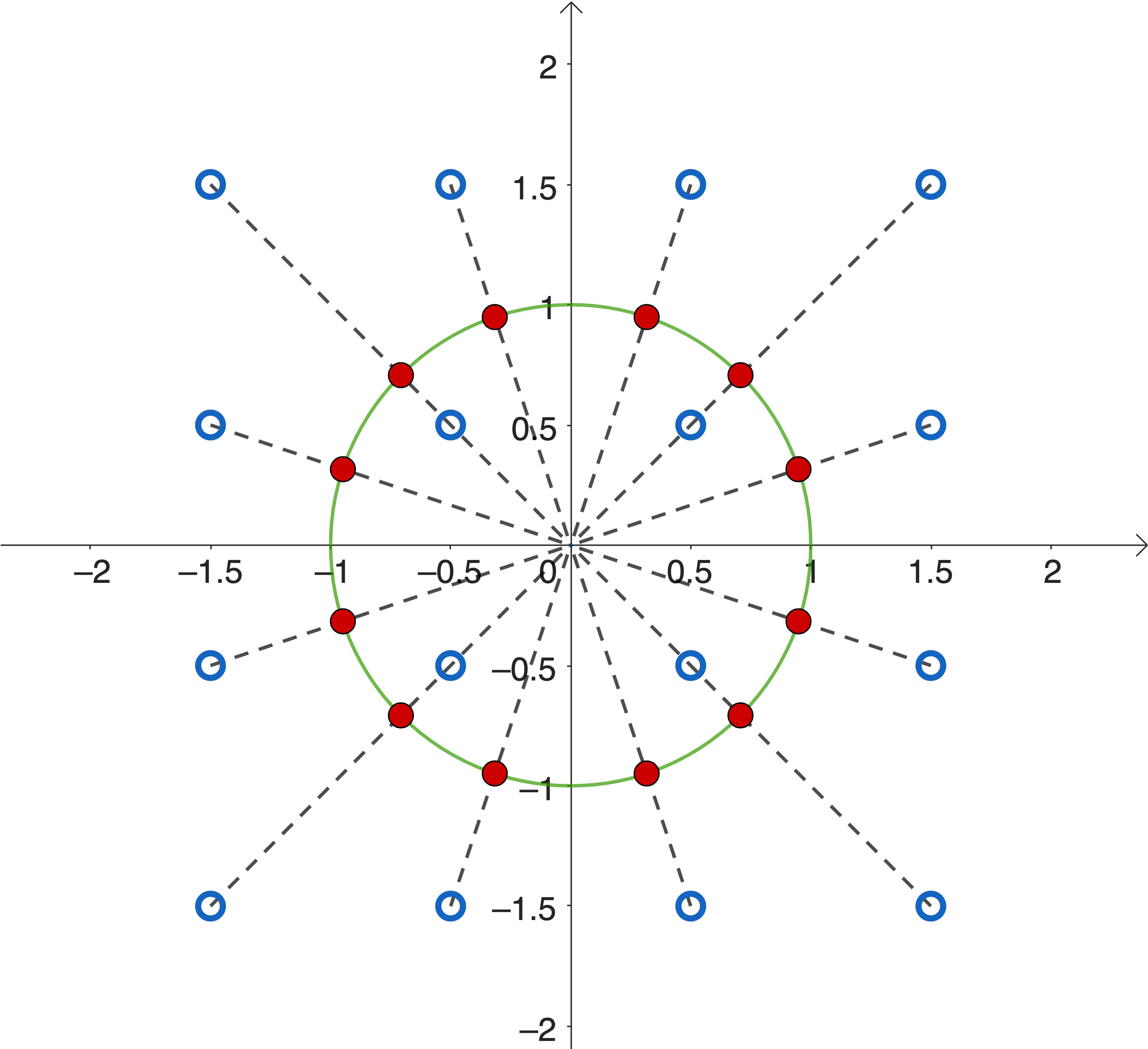}
    \end{subfigure} 
    \vspace{-2mm}
    \caption{This figure illustrates the quantization codebook of our method when $B=2$ in the 2-dimensional space. The empty blue points in the left panel represent the set $\mathcal{G}$, i.e., a set of vectors on a uniform grid. The solid red points in the right panel represent the normalized vectors in $\mathcal{G}$. Applying a random rotation on the red points yields the codebook $\mathcal{G}_r$.}
    \label{fig:codebook}
    \vspace{-4mm}
\end{figure}

\subsubsection{Constructing a Codebook} We notice that the simple extension in \cite{rabitq} provides sub-optimal accuracy because it ignores that the vectors can be hosted in a $D$-dimensional space. 
In contrast, it pads these vectors to $(B\cdot D)$ dimensions and applies RaBitQ in the $(B\cdot D)$-dimensional space.
The codebook has the size of $2^{B\cdot D}$ and thus, it produces the quantization codes of $(B\cdot D)$ bits. 
In intuition, given the same number of bits, a quantization algorithm would produce larger error when the vectors have higher dimensionality itself.
Thus, it is likely that the sub-optimal accuracy is caused by the padding operation. 

Instead of padding the vectors and constructing a codebook with $2^{B\cdot D}$ vectors in the $(B\cdot D)$-dimensional space, we consider constructing a codebook with $2^{B\cdot D}$ vectors in the $D$-dimensional space. 
Nevertheless, to inherit the unbiasedness and error bound of the estimator in RaBitQ and support efficient computation at the same time, the construction of the codebook is highly restrictive. 
\underline{\textbf{First}}, the unbiasedness and the error bound (Lemma~\ref{lemma: restate}) hold on condition that the codebook is constructed by randomly rotating a set of \textit{unit} vectors. 
\underline{\textbf{Second}}, for efficient computation, it should better support to directly compute the inner product based on the quantization codes without an exhaustive decompression step. 
We propose the following codebook $\mathcal{G}_r$ to meet both requirements. 
\begin{align}
\mathcal{G} &:=\left\{ -\frac{2^{B}-1}{2} + u\ \Big|\ u=0,1,2,3,...,2^B-1\right\}^D \label{equ:shift}
\\ \mathcal{G}_r &:=\left\{  P\frac{\mathbf{y}}{\| \mathbf{y}\|}\  \Big|\  \mathbf{y} \in \mathcal{G} \right\} \label{equ:rotation}
\end{align}
Specifically, as is illustrated in Figure~\ref{fig:codebook}, we first consider the set of vectors on the uniform grid $\mathcal{G}$. Then we adopt their normalized vectors (i.e., $\frac{\mathbf{y}}{\| \mathbf{y}\|}$ in Equation~(\ref{equ:rotation})) and randomly rotate them by multiplying them with a random orthogonal matrix $P$) to form the codebook $\mathcal{G}_{r}$. 
The rationale is that (1) $\mathcal{G}_r$ is composed of randomly rotated unit vectors. Thus, it inherits the estimator in Lemma~\ref{lemma: restate} along with its unbiasedness and error bound~\footnote{For the new codebook, Lemma~\ref{lemma: restate} exactly holds. 
However, it is worth noting that the inner product between the data vector and the quantized data vector in the new codebook increases with respect to $B$. Thus, the error of the estimator decreases.}; and (2) the vectors are generated by shifting, normalizing and rotating the vectors that consist of $B$-bit unsigned integers (Equation~(\ref{equ:shift}) and (\ref{equ:rotation})). This enables efficient computation of inner product (see Section~\ref{subsec: computing the estimator} for details).

In practice, the construction of the codebook is very simple. We only need to sample a random transformation matrix $P$ as RaBitQ does~\cite{rabitq}, and then the codebook $\mathcal{G}_r$ is determined. 
To store the codebook $\mathcal{G}_{r}$, we only need to store the sampled $P$, i.e., we maintain the codebook $\mathcal{G}_{r}$, which contains $2^{B\cdot D}$ vectors, conceptually. 
In particular, it is worth noting that when setting $B=1$, the codebook is exactly the same as that of the original RaBitQ.

\subsubsection{Computing the Quantization Codes of Data Vectors}
\label{subsubsec: exact quantization code}
Next we find for each data vector its nearest vector $\mathbf{\bar o}$ in the codebook $\mathcal{G}_r$ as its quantized vector. Formally, for a vector $\mathbf{o}$, we target to find $\mathbf{\bar o} \in \mathcal{G}_r$ such that $\| \mathbf{\bar o}-\mathbf{o}\|^2$ is minimized. 
Let $\mathbf{\bar y}$ be the corresponding vector of $\mathbf{\bar o}$ in $\mathcal{G}$, i.e., $\mathbf{\bar o}=P\mathbf{\bar y}/\|\mathbf{\bar y}\|$.
Following \cite{rabitq}, we simplify the problem as follows.
\begin{align}
    \mathbf{\bar y} &=\argmin_{\mathbf{y} \in \mathcal{G}} \left\| P \frac{\mathbf{ y}}{\|\mathbf{y}\|}-\mathbf{o}\right\|^2 
    =\argmin_{\mathbf{y} \in \mathcal{G}} \left( 2 - 2\left<P \frac{\mathbf{ y}}{\|\mathbf{y}\|}, \mathbf{o} \right>\right) \label{eq 3.2: definition}
    \\ &= \argmax_{\mathbf{y} \in \mathcal{G}} \left<P \frac{\mathbf{ y}}{\|\mathbf{y}\|}, \mathbf{o} \right> 
    = \argmax_{\mathbf{y} \in \mathcal{G}} \left<\frac{\mathbf{ y}}{\|\mathbf{y}\|}, P^{-1}\mathbf{o} \right> \label{eq 3.2: optimziation}
\end{align}
where Equation (\ref{eq 3.2: definition}) is derived from the definition of $\mathbf{\bar y}$; Equation (\ref{eq 3.2: optimziation}) applies an orthonormal matrix $P^{-1}$ to both sides of the inner product.
For conciseness, let us denote $\mathbf{o}':=P^{-1} \mathbf{o}$. Now the question reduces to one of finding $\mathbf{\bar y}\in \mathcal{G}$ such that Equation (\ref{eq 3.2: optimziation}) is maximized.

A natural idea is to enumerate the vectors in $\mathcal{G}$, compute the value of Equation (\ref{eq 3.2: optimziation}) for each vector and find the vector $\mathbf{\bar y}$. However, enumerating all the vectors in $\mathcal{G}$ is not feasible as $\mathcal{G}$ contains $2^{B\cdot D}$ vectors. 
We observe that the enumeration can be significantly pruned based on the following lemma. The proof is left in Appendix~\ref{appendix: lemma 3.1} due to page limit.  

\begin{lemma}
\label{lemma: quantization codes}
    Let $\mathbf{\bar y} =\argmax_{\mathbf{y} \in \mathcal{G}} \left<\frac{\mathbf{ y}}{\|\mathbf{y}\|}, \mathbf{o}' \right>$. Then $\exists t > 0$ such that $\forall \mathbf{y} \in \mathcal{G}, \| t\cdot \mathbf{o}' - \mathbf{\bar y}\| \le \| t\cdot \mathbf{o}' - \mathbf{y}\|$.  
\end{lemma}
In intuition, this lemma indicates that for a vector $\mathbf{o}'$, if a vector $\mathbf{\bar y}$ has the largest \textit{cosine similarity} 
from $\mathbf{o}'$ among the set $\mathcal{G}$, then there must be a re-scaling factor $t$ such that $\mathbf{\bar y}$ also has the smallest \textit{Euclidean distance} from the re-scaled vector $t\cdot \mathbf{o}'$ among the set $\mathcal{G}$.
Note that $\mathcal{G}$ is a set of vectors on uniform grids, the nearest vector of $t\cdot \mathbf{o}'$ in $\mathcal{G}$ can be easily computed by rounding. 
This is to say, if we ``enumerate every re-scaling factor $t$'' and collect all the vectors produced by rounding $t\cdot \mathbf{o}'$, then the target vector $\mathbf{\bar y}$ must be included in the vectors that are collected in the process. 
In practice, we do not need to enumerate every real value $t>0$ because when two re-scaling factors are extremely close to each other, they would produce exactly the same vector after rounding. 
Thus, we only need to enumerate the critical values which change the results of rounding. 
Specifically, for a given $t$, assume that $t\cdot \mathbf{o}'[i]$ is currently rounded to a number $x$. Then the next critical value of $t$ that rounds $t\cdot \mathbf{o}'[i]$ to $(x+1)$ is $(x+0.5)/\mathbf{o}'[i]$.
As there are $D$ dimensions and each dimension has $2^{B-1}$ different values after rounding~\footnote{It is $2^{B-1}$ instead of $2^B$ because $\mathbf{\bar y}$ is in the same orthant of $\mathbf{o}'$.}, we will in total enumerate up to $( D\cdot 2^{B-1}) $ critical values and correspondingly, up to $( D\cdot 2^{B-1}) $ vectors in $\mathcal{G}$. 

To efficiently implement this idea, we enumerate the critical values in an ascending order. The algorithm is presented in Algorithm~\ref{code: quantization code}. 
In this process, we dynamically maintain the vector produced by rounding with $\mathbf{y}_{cur}$ (the vector that is currently being enumerated).
We use two variables to maintain $\left< \mathbf{y}_{cur}, \mathbf{o}'\right>$ and $\| \mathbf{y}_{cur}\|$~\footnote{For $\mathbf{y}_{cur}$, we use two variables to store $\left< \mathbf{y}_{cur}, \mathbf{o}'\right>$ and $\| \mathbf{y}_{cur}\|$. Whenever $\mathbf{y}_{cur}$ is updated, we can update these variables to obtain the new values of $\left< \mathbf{y}_{cur}, \mathbf{o}'\right>$ and $\| \mathbf{y}_{cur}\|$ efficiently.}.
$v_{max}$ maintains the maximum value of $\left< \mathbf{y}_{cur}/\| \mathbf{y}_{cur}\|, \mathbf{o}'\right>$ based on the vectors that have been enumerated so far. $t_{max}$ maintains the re-scaling factor which produces the maximum value $v_{max}$.
The enumeration starts from $t=0$ (line 1-2).
Then iteratively, we enumerate the next smallest critical value (line 3-4). 
Based on the new re-scaling factor $t$, the vector $\mathbf{y}_{cur}$ will change in only one dimension. 
Thus, we can update $\mathbf{y}_{cur}, \left< \mathbf{y}_{cur}, \mathbf{o}'\right>$ and $\| \mathbf{y}_{cur}\|$ in $O(1)$ time (line 5). 
During the enumeration, we record the re-scaling factor which produces the maximum $\left<\mathbf{y}_{cur}/\|\mathbf{y}_{cur}\|, \mathbf{o}'\right>$ (line 6-7).
The enumeration terminates when all the critical values have been enumerated (line 3).
Finally, based on the re-scaling factor $t_{max}$, we find the $\mathbf{\bar y}$ via re-scaling and rounding (line 8). 
We represent $\mathbf{\bar y}$ and store it as a vector of unsigned integers $\mathbf{\bar y}_u$. Specifically, we have $\mathbf{\bar y}_u=\mathbf{\bar y} + (2^B-1)/2\cdot \mathbf{1}_D$ where $\mathbf{1}_D$ is the vector whose coordinates are all ones (line 9).

The overall time complexity of the algorithm is $O(2^B\cdot D\log D)$ because in total we enumerate $(2^{B-1}\cdot D)$ critical values and a min-heap is needed to find the next smallest critical value during the process (its maintenance takes $O(\log D)$ time).
We note that this time complexity is good enough for practical usage as our algorithm is designed for vector compression, i.e., $B$ is small. 
According to our experimental studies in Section~\ref{subsubsec: ANN}, $B=7$ suffices to stably produce $>99\%$ recall and $B=5$ suffices to stably produce $>95\%$ recall. 
Under these settings, the quantization of a million-scale dataset of 3,072 dimensions can finish in a few minutes.

\begin{algorithm}[tbh]
\DontPrintSemicolon
\SetKwData{and}{and}
\SetKwInOut{Input}{Input}\SetKwInOut{Output}{Output}
\Input{A $D$-dimensional vector $\mathbf{o}'$; the number of bits per dimension $B$.}
\Output{The quantization code $\mathbf{\bar y}_u$.}
\BlankLine
$t \leftarrow 0, v_{max}\leftarrow 0, t_{max} \leftarrow 0$\;
Initialize $\mathbf{y}_{cur}, \left< \mathbf{y}_{cur}, \mathbf{o}'\right>$ and $\| \mathbf{y}_{cur}\|$ with $t=0$\;
\While {some critical values have not been enumerated}{
Update $t$ with the next smallest critical value\;
Update $\mathbf{y}_{cur}, \left< \mathbf{y}_{cur}, \mathbf{o}'\right>$ and $\| \mathbf{y}_{cur}\|$ with the new $t$\;
\If{$\left< \mathbf{y}_{cur}, \mathbf{o}'\right>/\| \mathbf{y}_{cur}\| > v_{max}$}{
    $v_{max} \leftarrow \left< \mathbf{y}_{cur}, \mathbf{o}'\right>/\| \mathbf{y}_{cur}\|, t_{max}\leftarrow t$\;
}
}
Compute $\mathbf{\bar y}$ via re-scaling and rounding $\mathbf{o}'$ with $t_{max}$\;
\textbf{return} $\mathbf{\bar y}_u$ where $\mathbf{\bar y}_u=\mathbf{\bar y} + (2^B-1)/2\cdot \mathbf{1}$\;
\caption{\textbf{Quantize}}
\label{code: quantization code}
\end{algorithm}

\subsection{Computing the Estimator}
\label{subsec: computing the estimator}
Recall that we target to estimate the inner product $\left<\mathbf{o}, \mathbf{q} \right>$ to further estimate the squared distances (Section~\ref{subsec: pre-rabitq}). We adopt the estimator of RaBitQ to inherit its unbiasedness and error bound~\footnote{This conclusion can be directly yielded from the proof in the original RaBitQ paper~\cite{rabitq}, i.e., Lemma~\ref{lemma: restate} holds if (1) the codebook is a set of randomly rotated unit vectors; and (2) $\mathbf{\bar o}$ is the nearest vector of $\mathbf{o}$ in the codebook.} (see empirical verification in Section~\ref{subsubsec: space-accuracy}), i.e., we use $\left< \mathbf{\bar o}, \mathbf{q}\right>/\left< \mathbf{\bar o}, \mathbf{o} \right>$ to estimate $\left< \mathbf{o,q}\right>$. The denominator $\left< \mathbf{\bar o}, \mathbf{o} \right>$ is only related to the data vector and its quantized vector, so it can be pre-computed in the index phase. Thus, we only need to compute $\left< \mathbf{\bar o}, \mathbf{q}\right>$. Recall that $\mathbf{\bar o}=P \frac{\mathbf{\bar y}}{\|\mathbf{\bar y}\|}$. The following equations illustrate how it can be computed.
\begin{align}
    \left< \mathbf{\bar o}, \mathbf{q}\right>&=\left< P \frac{\mathbf{\bar y}}{\|\mathbf{\bar y}\|}, \mathbf{q}\right>=\left< \frac{\mathbf{\bar y}}{\|\mathbf{\bar y}\|}, P^{-1}\mathbf{q}\right> =\frac{1}{\|\mathbf{\bar y}\|}\left< \mathbf{\bar y}, \mathbf{q}'\right>\label{eq 3.3: reverse}
    \\&=\frac{1}{\|\mathbf{\bar y}\|}\left( \left< \mathbf{\bar y}_u, \mathbf{q}'\right>-\frac{2^B-1}{2} \sum_{i=1}^{D} \mathbf{q}'[i] \right) \label{eq 3.3: represent}
\end{align}
Here $\mathbf{q}'$ denotes $P^{-1}\mathbf{q}$; Equation (\ref{eq 3.3: reverse}) applies an orthonormal matrix $P^{-1}$ to both sides of the inner product;
and Equation (\ref{eq 3.3: represent}) expresses $\mathbf{\bar y}$ with its quantization code $\mathbf{\bar y}_u$, i.e., $\mathbf{\bar y}=\mathbf{\bar y}_u - (2^B-1)/2\cdot \mathbf{1}_D$.

Note that $\| \mathbf{\bar y}\|$ is only related to the quantized vectors and can be pre-computed in the index phase. $\sum_{i=1}^{D} \mathbf{q}'[i]$ is only related to the query vector. It can be computed once and its time costs can be shared by many data vectors. Thus, the remaining task is the computation of $\left< \mathbf{\bar y}_u, \mathbf{q}'\right>$, i.e., the inner product between a vector of unsigned integers and a vector of floating-point numbers.
When $B=1$ (the case of the original RaBitQ), RaBitQ's implementation can be directly applied~\cite{rabitq}.
When $B$ equals to 4 or 8, the implementations in existing systems (for computing the inner product between a vector of 4-bit or 8-bit unsigned integers and a vector of floating-point numbers) can be directly  applied~\cite{blink_scalar_quantization_graph,douze2024faisslibrary}.
Other settings of $B$'s can be implemented by splitting 
a vector of $B$-bit unsigned integers into several parts, where each part has the size of the power of 2 (e.g., a vector of 9-bit unsigned integers can be split into a binary vector and a vector of 8-bit unsigned integers). We will discuss the details of the idea later in Section~\ref{subsec: distance comparison with rabitq}.

\subsection{Summary and Theoretical Analysis}
\label{subsec: summary}
We summarize the workflow of the extended RaBitQ as follows. In the index phase, the algorithm constructs a quantization codebook by sampling a random orthogonal matrix $P$. It then applies $P^{-1}$ to the data vectors, normalizes them to obtain $\mathbf{o}'$ and computes their quantization codes via Algorithm~\ref{code: quantization code}. 
In the query phase, when a query comes, it first applies $P^{-1}$ to the query vector and normalizes it to obtain $\mathbf{q}'$. It can then unbiasedly estimate squared distances based on Equation (\ref{eq 3.3: represent}) and Equation (\ref{eq:dis to ip}).

Recall that as has been proved in \cite{2017_focs_additive_error} and discussed in Section~\ref{subsec: challenges}, 
to achieve an error bound $\epsilon$ where $\frac{1}{\epsilon^2} \log \frac{1}{\delta} > D$, the minimum required number of bits is $\Theta \left(D\log \left( \frac{1}{D}\cdot \frac{1}{\epsilon^2 } \log \frac{1}{\delta}\right) \right)$. The following theorem presents that our method achieves this asymptotic optimality.
The proof is left in Appendix~\ref{appendix: proof main theorem} due to page limit.
\begin{theorem}
\label{theorem: main}
For $\epsilon>0$ where $\frac{1}{\epsilon^2}\log \frac{1}{\delta} > D$,
to ensure that 
the error of the estimator is bounded by $\epsilon$ with the probability of at least $1-\delta$, the algorithm requires $B=\Theta \left(\log \left( \frac{1}{D}\cdot \frac{1}{\epsilon^2 } \log \frac{1}{\delta}\right) \right)$.
\end{theorem}
It is worth noting that the number of bits needed for each dimension $B$ is \textit{logarithmic} wrt $\epsilon^{-2}$ and is \textit{negatively} related to the dimensionality $D$. 
Besides the asymptotic analysis, 
to provide a more quantitative reference for practitioners, we would also like to present an empirical formula about the error. Note that when $B=1$ (the original setting of RaBitQ~\cite{rabitq}), the error is bounded by $O(1/\sqrt{D})$ with high probability. When a larger $B$ is used, the error is supposed to decay exponentially. Thus, we present the empirical formula in the following form.

\begin{remark}[Empirical Formula]
Let $\epsilon$ be the absolute error of estimating inner product of unit vectors. With >99.9\% probability, we have $\epsilon < 2^{-B} \cdot c_{\epsilon }/\sqrt{D}$ where $c_{\epsilon}=5.75$.
\end{remark}

We note that the constant in this empirical formula is measured with experimental studies, i.e., we use our algorithm to estimate inner product and collect the statistics of errors (see Section~\ref{subsubsec: measure constants}).

\smallskip
\noindent\textbf{Comparison with the Algorithmic Proof in \cite{2017_focs_additive_error}.}
In the theoretical study~\cite{2017_focs_additive_error}, there is an algorithmic proof which achieves the asymptotic optimality. 
However, we note that the algorithmic proof is less practically applicable.
Specifically, this proof relies on the operation which represents different integers with different number of bits.
Based on this operation, \cite{2017_focs_additive_error} proves that the total number of bits needed is asymptotically optimal.
This algorithm is unfriendly to real-world systems because it is unclear how a sequence of integers represented by different number of bits can be stored in alignment with each other. 
In contrast, our method achieves both the asymptotic optimality and the practicality. 
To the best of our knowledge, our study is also the first which achieves both desiderata at the same time.
We would like to emphasize that our work does not target the improvement in terms of theory. Instead, we propose a practically applicable algorithm and prove that it is asymptotically optimal.

\section{Applying The Extended RaBitQ to in-memory ANN}
\label{sec:ExRaBitQ for ANN}

\subsection{Using the Extended RaBitQ with IVF}

Next we apply the extended RaBitQ method to the in-memory ANN. 
Recall that as is discussed in Section~\ref{sec: introduction}, the present study targets to compress vectors with a moderate compression rate such that it can avoid storing raw vectors in RAM while still producing good overall recall (e.g., >90\%, >95\% or >99\%).
Scalar quantization (SQ) and its variants are the state-of-the-art and also the most popular methods in real-world systems for this  need~\cite{douze2024faisslibrary,milvus,blink_scalar_quantization_graph,survey_vector_database_2024}.
We note that these methods are usually used together with  IVF~\cite{douze2024faisslibrary,milvus,survey_vector_database_2024} or graph-based indices~\cite{blink_scalar_quantization_graph} for in-memory ANN query.
In particular, IVF is a popular method that has been widely deployed in real-world systems for vector search due to its simplicity and effectiveness~\cite{douze2024faisslibrary,milvus,survey_vector_database_2024}.
It has tiny index size and can be easily combined with various quantization methods due to its sequential memory access pattern in querying.
For graph-based indices, we note that due to their random memory access pattern of querying, combining them with SQ entails highly non-trivial efforts in engineering optimization to make the combined method work competitively~\cite{blink_scalar_quantization_graph}.
Thus, in this paper, we focus on using the extended RaBitQ method together with the IVF index and leave its combination with the graph-based indices as future work.

Specifically, the IVF method partitions the set of data vectors into many clusters (e.g., via KMeans) in the index phase. Then when a query comes, it finds a few nearest centroids of the clusters and considers the vectors in these clusters as candidates of ANN. 
When IVF is used in combination with SQ, it computes an estimated distance for every candidate based on its quantization code and then returns the vector with the smallest estimated distance as the NN.
Note that for reducing the memory consumption, when IVF is used with SQ, it does not store the raw vectors in RAM and thus, there is no re-ranking based on the raw vectors.
When using the same number of bits for quantization, our method can produce better accuracy than SQ and its variants (see experimental results in Section~\ref{subsubsec: space-accuracy}). 
In addition, its computation can be conducted with exactly the same implementations of SQ (Section~\ref{subsec: computing the estimator}).
Thus, it is expected that simply replacing SQ with our method suffices to produce better time-accuracy trade-off (i.e., it improves the accuracy while maintaining the efficiency). 

\begin{figure}[thb]
    \vspace{-2mm}
    \centering
    \includegraphics[width=\linewidth]{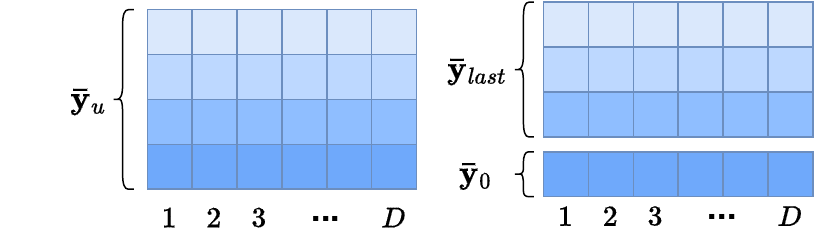}
    \vspace{-4mm}
    \caption{Decomposition of the Quantization Code $\mathbf{\bar y}_u$.}
    \label{fig:decomposition}
    \vspace{-4mm}
\end{figure}
\subsection{Conducting Distance Comparison with the Extended RaBitQ}
\label{subsec: distance comparison with rabitq}
Beyond trivially replacing SQ with our extended RaBitQ, we note that the efficiency of the method can be further improved.
In particular, a recent study finds that to reliably find the NN from a set of candidates, it is not necessary to compute an exact distance for every candidate~\cite{adsampling}.
If an estimated distance can confirm that a candidate is unlikely to be the NN (e.g., a lower bound of the estimated distance is greater than the distance of the NN that has been searched so far), then the candidate can be discarded.
Otherwise, it can incrementally compute a more accurate estimated distance until (1) it can confirm that a candidate is unlikely to be the NN; or (2) the exact distance is computed~\cite{adsampling}.

Based on the idea above, in our case, we note that it is not always necessary to estimate a highly accurate distance with all bits of a quantization code. Instead, we first use a subset of bits of the quantization code to efficiently estimate a distance with lower accuracy. If this estimated distance can confirm that a candidate is unlikely to be the NN, then we discard it. Otherwise, we access the remaining bits to compute a high-accuracy estimated distance based on the full bits of the quantization code. 

Specifically, in the quantization code $\mathbf{\bar y}_u$, every dimension corresponds to an unsigned integer of $B$ bits. We observe that the concatenation of the most significant bits of all the dimensions of $\mathbf{\bar y}_u$ (see Figure~\ref{fig:decomposition}) exactly equals to the quantization code $\mathbf{\bar x}_b$ of the original RaBitQ. 
This is because both of them are decided by the orthant of the vector $\mathbf{o}'$.
This motivates us to split a quantization code $\mathbf{\bar y}_u$ into two parts, the code of the most significant bits $\mathbf{\bar y}_{0}$ and the code of the remaining bits $\mathbf{\bar y}_{last}$, where $\mathbf{\bar y}_u=2^{B-1}\cdot \mathbf{\bar y}_{0} + \mathbf{\bar y}_{last}$ and $\mathbf{\bar y}_{0}=\mathbf{\bar x}_{b}$.
In the query phase, for a set of candidates, we first estimate their distances based on their $\mathbf{\bar y}_{0}$'s, i.e., we compute $\left<\mathbf{\bar y}_{0}, \mathbf{q}' \right>$ with FastScan~\cite{fastscanavx2} as the original RaBitQ does, and further compute the estimated distances based on Lemma~\ref{lemma: restate} and Equation (\ref{eq:dis to ip}).
Note that this estimated distance is exactly the estimated distance produced by the original RaBitQ and it has a theoretical error bound. 
Then we use the bound to decide whether the candidate is unlikely to be the NN.
If so, we drop the candidate. Otherwise, we access the code of the remaining bits $\mathbf{\bar y}_{last}$ to incrementally compute an estimated distance based on the full bits as follows.
\begin{align}
    \left< \mathbf{\bar y}_u,\mathbf{q}'\right>&=\left< 2^{B-1}\cdot \mathbf{\bar y}_0 + \mathbf{\bar y}_{last},\mathbf{q}'\right> \label{eq 4.2: decomposition}
    \\&= 2^{B-1} \cdot \left< \mathbf{\bar y}_0,\mathbf{q}'\right> + \left<\mathbf{\bar y}_{last}, \mathbf{q}'\right> \label{eq 4.2: simplify}
\end{align}
In particular, as $\left<\mathbf{\bar y}_{0}, \mathbf{q}' \right>$ has been computed and can be reused, we only need to access $\mathbf{\bar y}_{last}$ and compute $\left<\mathbf{\bar y}_{last}, \mathbf{q}' \right>$.
When $B=5$ and $B=9$, $\mathbf{\bar y}_{last}$ corresponds to vectors of 4-bit and 8-bit unsigned integers respectively and the computation of $\left<\mathbf{\bar y}_{last}, \mathbf{q}' \right>$ can be realized with existing implementations~\cite{douze2024faisslibrary,blink_scalar_quantization_graph,glass}.
Moreover, we provide the implementations (including designs in both compact storage and efficient computation of inner product) for more settings of $B$'s ($B=3,4,7,8$) as they also provide valuable trade-off among space, time and accuracy.
Due to the limit of space, we leave the details in our code repository.

\section{experiments}
\label{sec:experiments}

\subsection{Experimental Setup}
\label{subsec:experimental setup}

\smallskip\noindent\textbf{Experimental Platform.}
All experiments are run on a machine with two Intel Xeon Gold 6418H@4.0GHz CPUs (with Sapphire Rapids architecture, 48 cores/96 threads) and 1TB RAM. 
The C++ source codes are compiled by GCC 11.4.0 with \texttt{-Ofast -march=native} under Ubuntu 22.04 LTS. 
For all methods, by following most of the existing studies and benchmarks~\cite{fu2019fast, annbenchmark}, the search performance is evaluated in a single thread and the indexing time is measured using multiple threads.
The source code is available at \url{https://github.com/VectorDB-NTU/Extended-RaBitQ}.

\smallskip\noindent\textbf{Datasets.}
We evaluate the performance of different algorithms with six public real-world datasets with varying dimensionality and data types, whose details are presented in Table~\ref{tab:data}. 
These datasets encompass both (1) widely adopted benchmarks for evaluating ANN algorithms~\cite{li2019approximate,annbenchmark,rabitq} (such as Word2Vec, MSong and GIST~\footnote{\url{https://www.cse.cuhk.edu.hk/systems/hash/gqr/datasets.html.}}) and (2) embeddings generated by advanced models (including OpenAI-1536~\footnote{\url{https://huggingface.co/datasets/Qdrant/dbpedia-entities-openai3-text-embedding-3-large-1536-1M}}, OpenAI-3072~\footnote{\url{https://huggingface.co/datasets/Qdrant/dbpedia-entities-openai3-text-embedding-3-large-3072-1M}},
Youtube~\footnote{\url{https://research.google.com/youtube8m/download.html}}, and MSMARCO~\footnote{\url{https://huggingface.co/datasets/Cohere/msmarco-v2.1-embed-english-v3}}).
It is worth highlighting that OpenAI-1536 and OpenAI-3072 are produced by the most powerful embedding model \textit{text-embedding-3-large} of OpenAI published in early 2024~\cite{openaiembedding}. 
They are generated for evaluating ANN algorithms by Qdrant~\cite{Qdrant}.
As for the set of query vectors, for the datasets which provide the query set themselves, we adopt their query set for testing. 
For others, we exclude 1,000 vectors from each dataset and use them as the query vectors.

\begin{table}[h]
\caption{Dataset Statistics}
\vspace{-4mm}
\label{tab:data}
\centering
\begin{tabular}{c|cccc}
\hline
Dataset & Size      & $D$ &  Query Size & Data Type \\ \hline
MSong & 992,272 & 420      &  200      & Audio \\
Youtube  & 999,000 & 1,024       &  1,000     & Video     \\
OpenAI-1536  & 999,000 & 1,536       &  1,000     & Text     \\
OpenAI-3072  & 999,000 & 3,072       &  1,000     & Text  \\
Word2Vec & 1,000,000 & 300      &  1,000      & Text \\
GIST    & 1,000,000 & 960       &  1,000     & Image  \\
MSMARCO    & 113,520,750 & 1,024       &  1,677     & Text  \\
 \hline
\end{tabular}
\end{table}
\smallskip\noindent\textbf{Algorithms.}
In the experiments regarding the trade-off between the space and the accuracy of distance estimation (Section~\ref{subsubsec: space-accuracy}), we evaluate the performance of the following methods.
\textbf{\underline{(1) RaBitQ (ext)}} is the extended RaBitQ method proposed in the current study.
\textbf{\underline{(2) RaBitQ (pad)}} is the simple extension of RaBitQ mentioned in the original RaBitQ paper which uses more bits by padding the vectors with zeros~\cite{rabitq}. 
\textbf{\underline{(3) SQ}} is the classic uniform scalar quantization method. It has been widely deployed in real-world systems~\cite{survey_vector_database_2024,PASE,milvus, spann}.
Specifically, SQ first collects the minimum value $v_l$ and maximum value $v_r$ among all the coordinates of all the vectors.
Then, the method uniformly splits the range of values $[v_l,v_r]$ into $2^B-1$ segments. Each floating point number is then rounded to its nearest boundary of the segments and is represented and stored as a $B$-bit unsigned integer. \textbf{\underline{(4) LVQ}} is the latest variant of SQ~\cite{blink_scalar_quantization_graph}. 
Different from SQ which collects the smallest and largest values $v_l$ and $v_r$ among all the vectors and performs quantization based on this \emph{global} range of values $[v_l,v_r]$, LVQ collects the $v_l$ and $v_r$ for every \emph{individual} vector and performs quantization of a vector based on its specific range of values $[v_l,v_r]$. 
\textbf{\underline{(5) PQ}} and \textbf{\underline{(6) OPQ}} are popular quantization methods which are usually used with a large compression rate. They are widely deployed in real-world systems~\cite{survey_vector_database_2024,PASE,milvus}. Note that these methods have two settings $k=4$ or $k=8$ ($k$ is the number of bits allocated for quantizing each sub-vector in these methods; see details in their original papers~\cite{jegou2010product,ge2013optimized}).
Since $k=8$ produces consistently better space-accuracy trade-off than $k=4$, we report the results of PQ and OPQ with $k=8$.
It is worth noting that when using the same number of bits per dimension, SQ, RaBitQ (ext) and LVQ have almost the same efficiency in computing inner product and Euclidean distances.
However, as has been reported~\cite{blink_scalar_quantization_graph,rabitq}, PQ and OPQ have significantly worse efficiency since their computation relies on frequently looking up tables in RAM.
Recall that we target to compress vectors such that we do not need to access raw vectors for re-ranking while still producing reasonable recall.
Therefore, we focus on evaluating these methods in a moderate compression rate, i.e., the number of bits per dimension ranges from 1 to 10. 
When a larger compression rate is used, none of methods can stably produce over 90\% recall without re-ranking. 
When a smaller compression rate is used, it would be wasteful since 10 bits per dimension suffice to produce nearly perfect recall.
In addition, in this experiment, as a pre-procession for all the methods, we center the datasets with their global centroid~\footnote{With the centroid $\mathbf{c}$, the centering operation is to replace every data and query vector $\mathbf{a}$ with $\mathbf{a-c}$.}.
In the experiments regarding the ANN query (Section~\ref{subsubsec: ANN}), based on the experimental results in Section~\ref{subsubsec: space-accuracy}, we compare our method with the most competitive baseline LVQ~\cite{blink_scalar_quantization_graph}.
We combine our method and LVQ with the IVF index as is discussed in Section~\ref{sec:ExRaBitQ for ANN}. 
In this experiment, as a pre-procession for all the methods, we center every cluster in the IVF index with its local centroid.
All the methods are optimized with the SIMD instructions till AVX512.

\smallskip\noindent\textbf{Performance Metrics.}
In the experiments regarding the trade-off between the space and the accuracy of distance estimation (Section~\ref{subsubsec: space-accuracy}), we measure the accuracy with (1) the \emph{average relative error} and (2) the \emph{maximum relative error} on the estimated squared distances. 
We measure the space with the number of bits per dimension. 
Specifically, it sums up all the space consumption of a vector and divides it by the dimensionality. 
Note that for our method, it also covers the space for storing two floating-point numbers for every vector, i.e., $\|\mathbf{o}_r-\mathbf{c}\|$ and $1/ (\|\mathbf{y}_u\|\cdot \left< \mathbf{o},\mathbf{\bar o}\right>)$. 
In the experiments regarding the ANN query (Section~\ref{subsubsec: ANN}),
we adopt \emph{recall} and \emph{average distance ratio} for measuring the accuracy of ANN search. 
\emph{Recall} is the percentage of successfully retrieved true nearest neighbors.
\emph{Average distance ratio} is the average of the distance ratios of the retrieved nearest neighbors over the true nearest neighbors. 
These metrics are widely adopted to measure the accuracy of ANN algorithms~\cite{li2019approximate,annbenchmark,c2lsh,sun2014srs,huang2015query, SISAP_metric}. 
We adopt \emph{query per second (QPS)}, i.e., the number of queries a method can handle in a second, to measure the efficiency. It is widely adopted to measure the efficiency of ANN algorithms~\cite{li2019approximate, graphbenchmark, annbenchmark}.
Following \cite{li2019approximate, graphbenchmark, annbenchmark}, the \emph{query time} is evaluated in a single thread and the search is conducted for each query individually (instead of queries in a batch).
All the metrics are measured on every single query and averaged over the whole query set. 
We also report the time costs of different methods in the index phase. 

\smallskip\noindent\textbf{Parameter Settings in ANN Query.}
In the IVF index, we set the number of clusters to be 4,096 for million-scale datasets by following the suggestions of Faiss~\cite{faiss_github}. 
For the MSMARCO dataset, we use 262,144 clusters.
For our method, we implement our algorithm with $B=3, 4, 5, 7, 8, 9$ and conduct ANN query with the strategy presented in Section~\ref{sec:ExRaBitQ for ANN}. 
For the alignment of data, we pad the vectors with zeros such that their dimensionality is a multiple of 64 (i.e., we pad the dimensionality of MSong to 448 and that of Word2Vec to 320).
For LVQ, we apply the settings $B=4$ and $B=8$ in its original paper~\cite{blink_scalar_quantization_graph}.

\subsection{Experimental Results}
\label{subsec: experimental results}

\subsubsection{Space-Accuracy Trade-Off for Distance Estimation}
\label{subsubsec: space-accuracy}

\begin{figure*}[th]
  \centering 
    \includegraphics[width=\textwidth]{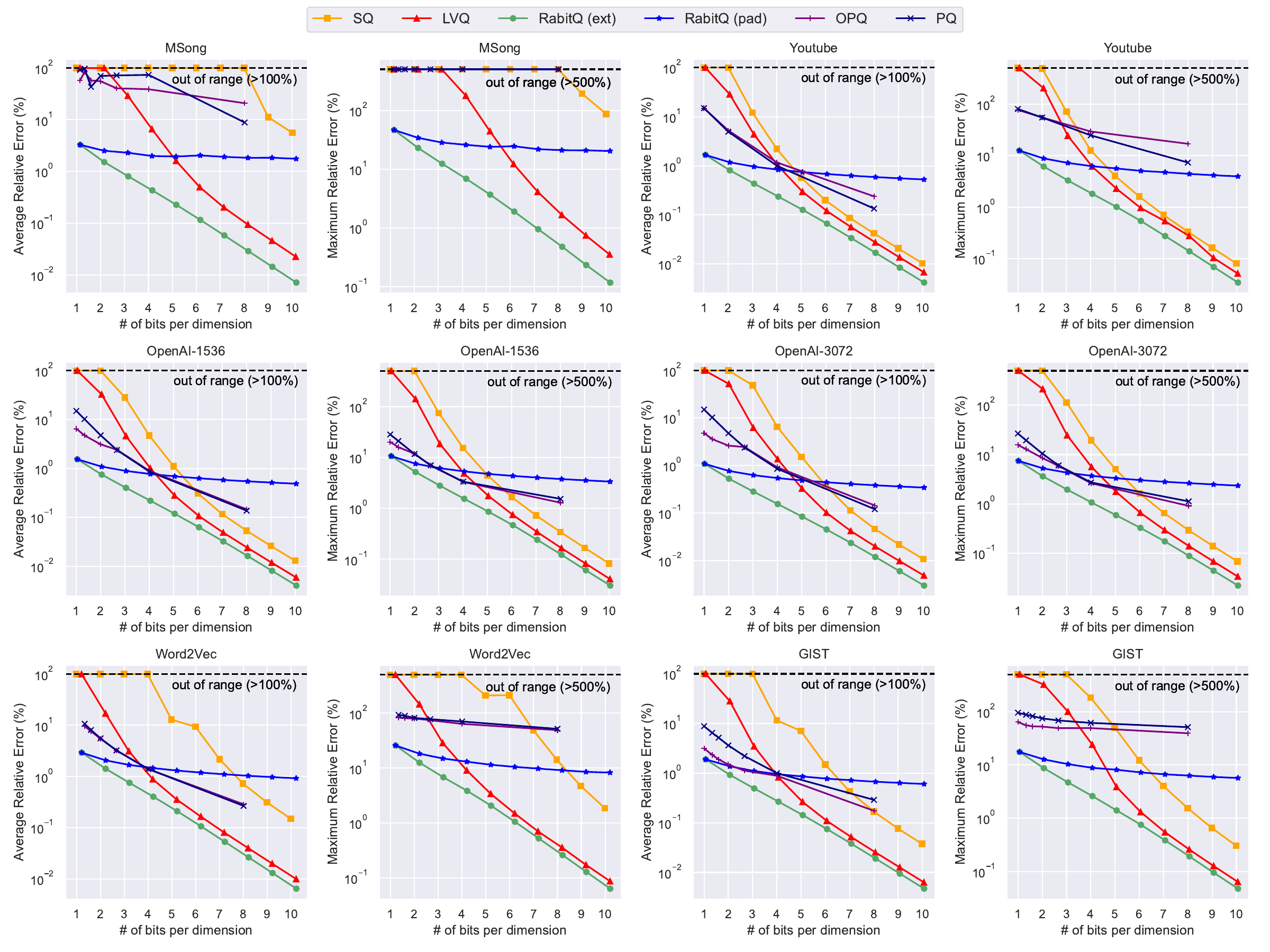}
  \vspace*{-6mm}
  \centering 
  \caption{Space-Accuracy Trade-Off for Distance Estimation (Log-Scale). }
  \vspace*{-4mm}
  \label{figure:space-accuracy extend rabitq}
\end{figure*}

In this experiment, we evaluate different quantization methods by using them for estimating the distances between data vectors and query vectors. 
For each method, in Figure~\ref{figure:space-accuracy extend rabitq}, we vary their number of bits and plot the curves of the average relative error (left panels, lower left is better) and maximum relative error (right panels, lower left is better) to investigate their space-accuracy trade-off.
We, in particular, focus on the number of bits from $B=1$ to $B=10$ which correspond to a moderate compression rate. Note that $B=8$ suffices to produce $>99\%$ recall (Section~\ref{subsubsec: ANN}). 

According to Figure~\ref{figure:space-accuracy extend rabitq}, our method (the green curve) stably achieves better accuracy than all the baseline methods on all the tested datasets when using the same number of bits. 
Specifically, we have the following observations. 
\underline{(1) RabitQ (ext) v.s. SQ and LVQ}: We observe that when $B>6$, under the same number of bits, the average relative error of LVQ is consistently larger than ours by 1.3x-3.1x (the gap of SQ's error from ours is even larger). When $B<6$, the gap is even larger. It is worth noting that when $B=1$ or $B=2$, LVQ and SQ hardly produce reasonable accuracy and their errors are larger than ours by orders of magnitude. 
Note that for our method, LVQ and SQ, the computation of distances or inner product can be implemented in almost identical ways (Section~\ref{subsec: computing the estimator}).
This result implies that simply replacing SQ and LVQ in existing systems with our method would stably improve the performance. Moreover, when a small $B$ is adopted, the improvement would be especially significant. 
This unique advantage enables our method to compute a fairly accurate distance based on the first bit of the quantization codes and prune many candidates, which improves the efficiency.
It will be reflected 
later in Section~\ref{subsubsec: ANN}.
\underline{(2) PQ and OPQ:} We find that PQ and OPQ have reasonable accuracy in most datasets when the number of bits per dimension is small (e.g., 1 or 2). However, when the number of bits increases, the errors of these methods do not decay as fast as our method, SQ or LVQ.
In particular, PQ and OPQ fail to outperform SQ and LVQ usually when the number of bits per dimension is $\ge 4$. This result has also been observed in the LVQ paper~\cite{blink_scalar_quantization_graph}.
Moreover, it has also been reported that the efficiency of PQ and OPQ is significantly worse than SQ and LVQ because for computing distances or inner product, they rely on looking up tables in RAM~\cite{blink_scalar_quantization_graph,fastscan}. 
When FastScan is applied to accelerate the computation (which requires to set $k=4$), the space-accuracy trade-off would be even worse~\cite{fastscanavx2}.
These results indicate that PQ and OPQ can hardly be competitive in compressing high-dimensional vectors with moderate compression rates. 
\underline{(3) RaBitQ (pad):} We find that RaBitQ (pad) is competitive when the number of bits per dimension is small (e.g., 1 or 2). However, the error of RaBitQ (pad) also decays slowly. This is because as has been theoretically proved in Lemma~\ref{lemma B.3}~\cite{rabitq}, its error decays in a trend of $O(1/\sqrt{B\cdot D})$.
As a comparison, the errors of our method, SQ and LVQ decay in an exponential trend with respect to $B$. 

\begin{figure*}[ht]
  \centering 
    \includegraphics[width=\textwidth]{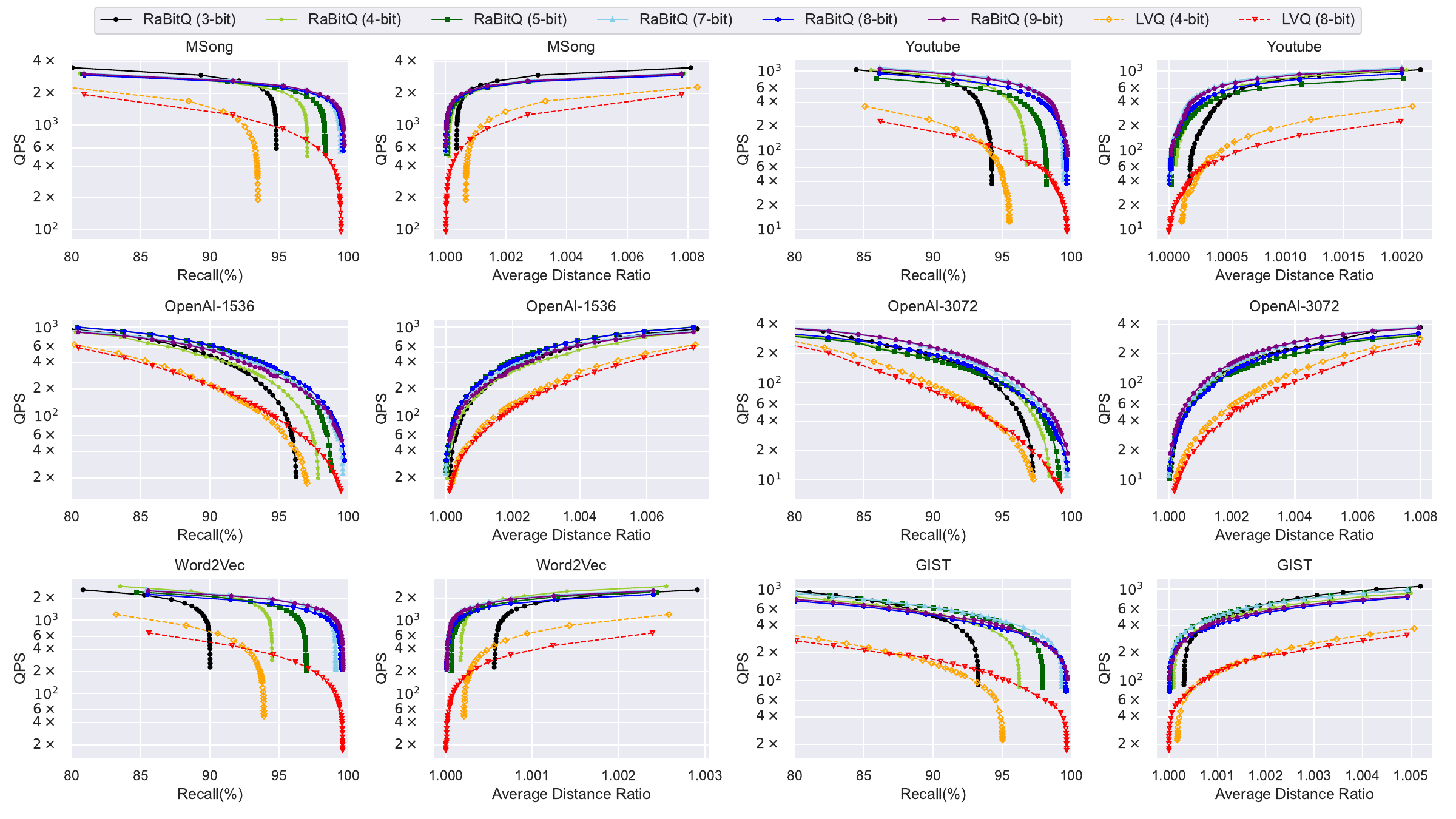}
  \vspace*{-6mm}
  \centering 
  \caption{Time-Accuracy Trade-Off for the ANN Query (Log-Scale), $K=100$. All the methods are combined with the IVF index. }
  \label{figure:time-accuracy extend rabitq}
\end{figure*}

\subsubsection{Time-Accuracy Trade-Off for ANN Query}
\label{subsubsec: ANN}

In this experiment, we evaluate different quantization methods by using them in combination with the IVF index for ANN queries. 
In Figure~\ref{figure:time-accuracy extend rabitq}, we plot for the methods their curves of ``QPS''-``Recall'' (left panels, upper right is better) and curves of ``QPS''-``Average Distance Ratio'' (right panels, upper left is better) by varying the number of clusters to probe to investigate their time-accuracy trade-off. In Table~\ref{tab: Space Consumption}, we report the space consumption of different methods for ANN query. We note that using the same $B$, our method has slightly larger space consumption (e.g., by less than 0.01 GB on million-scale datasets) than the baseline because our method uses FastScan to estimate distances batch by batch. When a batch is not full, we still allocate the memory for a full batch.

According to Figure~\ref{figure:time-accuracy extend rabitq}, we have the following observations. 
\underline{(1) Time-accuracy trade-off:} We observe that with the same number of bits (4-bit and 8-bit), our method outperforms LVQ in terms of time-accuracy trade-off. 
It is worth highlighting that the improvement in efficiency under the same recall is significant since our method can prune many candidates with the estimated distances based on the first bits of the quantization codes.
In addition, the estimation can be implemented with FastScan~\cite{fastscanavx2} which is highly efficient (Section~\ref{subsec: distance comparison with rabitq}).
\underline{(2) Recall:} We find that 4-bit, 5-bit and 7-bit quantization suffices to produce over $90$\%, $95$\% and $99$\% recall on all the tested datasets respectively without re-ranking. 
\underline{(3) Average distance ratio:} We observe that the average distance ratio produced by 5-bit quantization of our method is nearly perfect across all the datasets while it cannot achieve perfect recall (i.e., it usually produces $97\%$ recall only).
This indicates that the retrieved data vectors have their distances from the query extremely close to those of the true NNs. 
\underline{(4) Results on datasets from OpenAI:} We find that the ANN algorithm with our method on these datasets is highly robust, e.g., 3-bit quantization suffices to produce $>95\%$ recall. 

\begin{table*}[]
\caption{Space Consumption for ANN Query (GB).}
\label{tab: Space Consumption}
\vspace{-4mm}
\begin{tabular}{c|ccccccccc}
            & Raw Vectors   & RaBitQ-3 & RaBitQ-4 & RaBitQ-5 & RaBitQ-7 & RaBitQ-8 & RaBitQ-9 & LVQ-4 & LVQ-8 \\ \hline
MSong       & 1.56  & 0.18     & 0.23     & 0.28     & 0.39     & 0.44     & 0.49     & 0.22  & 0.41  \\
Youtube     & 3.82  & 0.40     & 0.52     & 0.64     & 0.87     & 0.99     & 1.11     & 0.51  & 0.98  \\
OpenAI-1537 & 5.72  & 0.59     & 0.77     & 0.95     & 1.31     & 1.48     & 1.66     & 0.75  & 1.47  \\
OpenAI-3072 & 11.44 & 1.19     & 1.54     & 1.90     & 2.62     & 2.97     & 3.33     & 1.49  & 2.92  \\
Word2Vec    & 1.12  & 0.13     & 0.17     & 0.21     & 0.28     & 0.32     & 0.35     & 0.16  & 0.30  \\
GIST        & 3.58  & 0.37     & 0.48     & 0.60     & 0.82     & 0.93     & 1.04     & 0.48  & 0.92 \\ 
MSMARCO        & 433.47  & 43.41     & 56.94     & 70.48     & 97.54     & 111.07     & 124.61     & 56.86  & 110.99 \\ \hline 
\end{tabular}
\end{table*}

\subsubsection{Time of Quantizing in the Indexing Phase}
\label{subsubsec: quantize time extended rabitq}

\begin{table}[]
\centering
\caption{Time for quantizing the data vectors in OpenAI-3072 ($\sim10^6$ vectors) with 96 threads/48 cores.}
\label{tab: quantization time extended rabitq}
\vspace{-4mm}
\begin{tabular}{cccccc}
\hline
$B$        & 1    & 2    & 3     & 4     & 5     \\
Time (s) & 43.8 & 47.7 & 52.6  & 58.1  & 64.6  \\ \hline
$B$        & 6    & 7    & 8     & 9     & 10    \\
Time (s) & 76.0 & 98.3 & 143.7 & 233.2 & 418.1 \\ \hline
\end{tabular}
\vspace{-2mm}
\end{table}

In this section, we report the time for quantizing the vectors in the index phase based on different $B$'s. This time includes both (1) the time for multiplying by $P^{-1}$ the data vectors and (2) the time for computing the quantization codes, i.e., Algorithm~\ref{code: quantization code}.
Due to the limit of space, we only report the results on the dataset OpenAI-3072 which has the highest dimensionality. It is clear that the quantization time for other datasets which have lower dimensionality would be smaller.
Based on the results in Figure~\ref{figure:time-accuracy extend rabitq} and Table~\ref{tab: quantization time extended rabitq}, we observe that when setting $B=5$ and $B=7$, it suffices to produce >$95\%$ and >$99\%$ recall respectively, and the quantization can finish in a few minutes. 
Thus, the time for quantizing data vectors in the index phase is not a bottleneck in practical usage. 
 
\subsubsection{Verifying the Scalability}
\begin{figure}[th]
  \centering 
    \includegraphics[width=\linewidth]{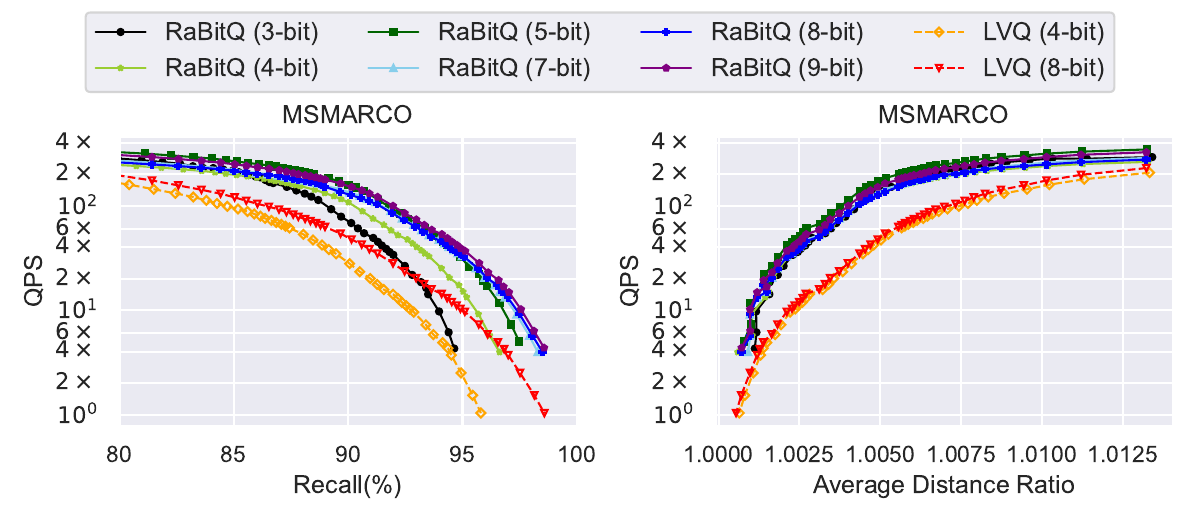}
  \vspace{-8mm}
  \caption{Verification Study for Scalability.}
  \vspace{-4mm}
  \label{figure:scalability}
\end{figure}
In this section, we study the scalability of our method on the MSMARCO dataset with about 100 millions of 1,024-dimensional data vectors. We report that the quantization of the dataset with $B=9$ can be finished in around 2 hours, which is not a bottleneck (as a comparison, building the IVF index for the dataset takes more than 1 day). Figure~\ref{figure:scalability} reports the time-accuracy trade-off for ANN queries. It shows that our method still achieves consistently better time-accuracy trade-off compared with LVQ. 
In particular, with our method, using $B=4$ suffices to produce >95\% recall without re-ranking. 
In this case, the space consumption of our method is 56.94 GB while the raw dataset takes 433.47 GB.

\subsubsection{Verifying the Unbiasedness}
\label{subsubsec: verify unbiasedness extended rabitq}
\begin{figure}[th]
  \centering 
    \includegraphics[width=\linewidth]
    {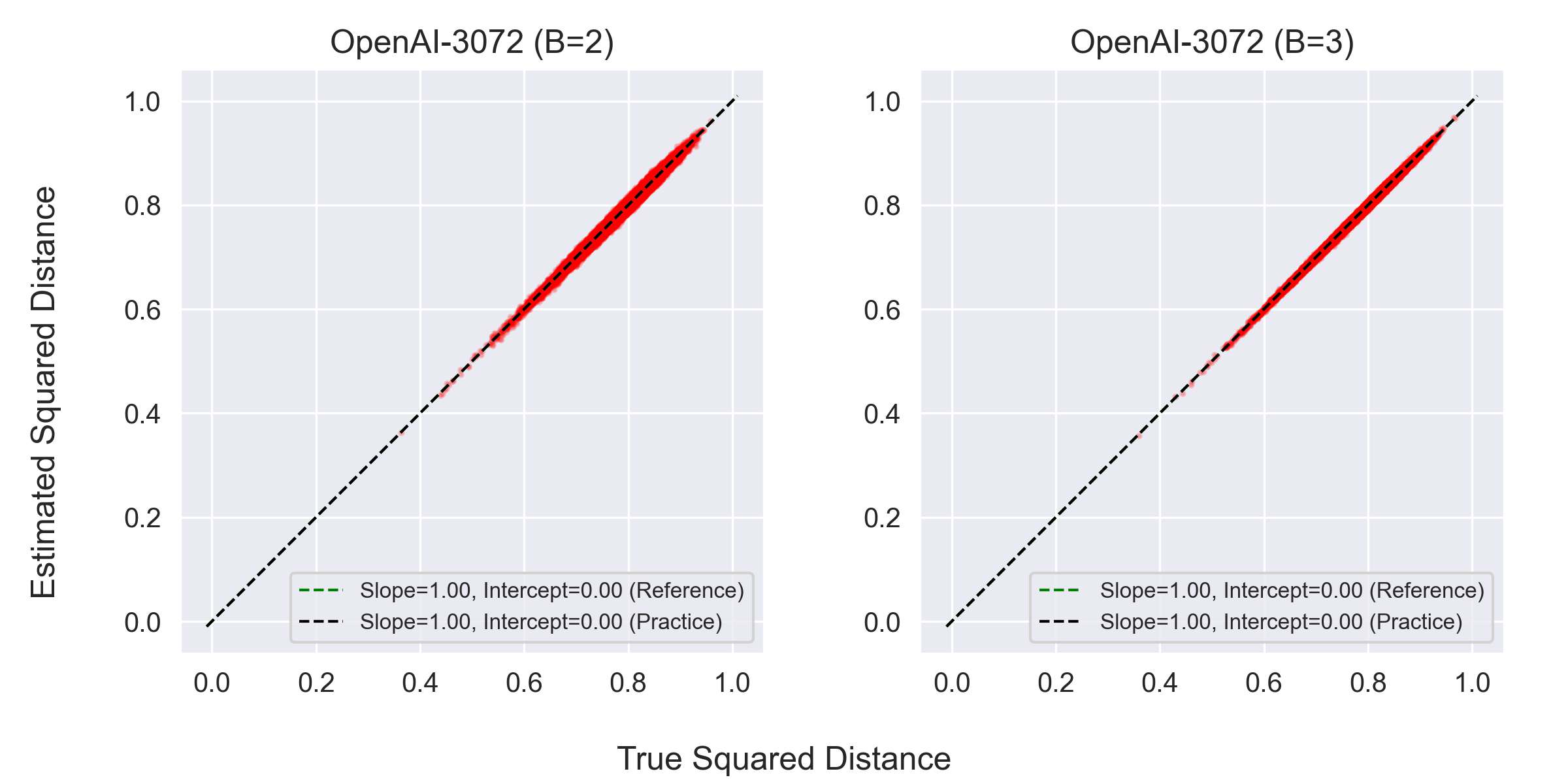}
  \vspace{-6mm}
  \caption{Verification Study for Unbiasedness.}
  \vspace{-4mm}
  \label{figure:unbiasedness extended rabitq}
\end{figure}

In this section, we verify that based on our new codebook with more vectors than those of the original RaBitQ method, the estimator is still unbiased. 
Due to the limit of space, we only present the results for $B=2,3$ in two panels of Figure~\ref{figure:unbiasedness extended rabitq} respectively.
In each panel, we collect around $10^7$ pairs of the estimated squared distances and the true squared distances (between the first 10 query vectors and about $10^6$ data vectors in the OpenAI-3072 dataset) which are normalized with the maximum true squared distances. 
To verify the unbiasedness, we fit these pairs with linear regression and plot the result with the black dashed line. 
Note that in all panels, this line has the slope of 1 and the y-axis intercept of 0.
This indicates that our method provides unbiased distance estimation.

\subsubsection{Measuring the Empirical Formula among $\epsilon$, $D$ and $B$}
\label{subsubsec: measure constants}
\begin{figure}[th]
  \centering 
    \vspace{-4mm}
    \includegraphics[width=\linewidth]
    {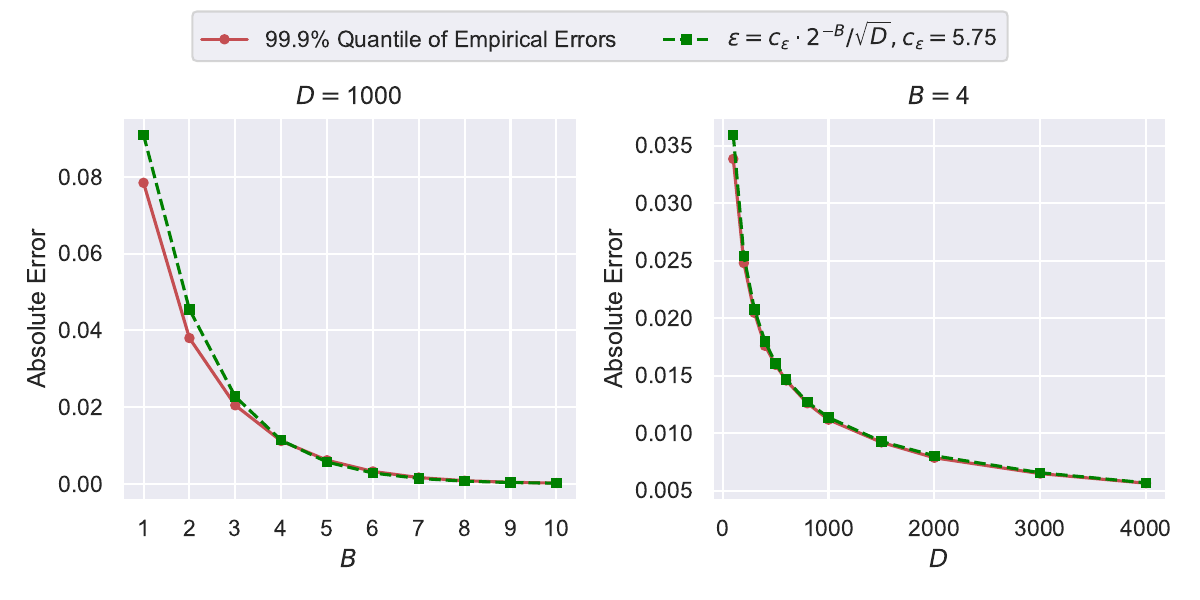}
  \vspace{-8mm}
  \caption{Measure the Constants in the Empirical Formula.}
  \vspace{-4mm}
  \label{figure:measure constants}
\end{figure}

In this section, we measure the constants in the empirical formula which presents the relationship among the absolute error of estimating inner product between unit vectors, $B$ and $D$ (see Section~\ref{subsec: summary}). In particular, for a pair of $B$ and $D$, we randomly sample $5\times 10^6$ pairs of unit vectors (i.e., each vector is sampled from standard Gaussian distribution and is normalized then), apply our method to estimate their inner product and collect the error of the estimation. 
In the left panel of Figure~\ref{figure:measure constants}, we fix $D=1000$ and plot the curve of the 99.9\% quantile of the empirical errors with respect to $B$ (the red curve). 
In the right panel, on the other hand, we fix $B=4$ and plot the curve of the 99.9\% quantile of the empirical errors with respect to $D$ (the red curve). 
Note that our empirical formula is in the form of $\epsilon<c_{\epsilon}\cdot 2^{-B}/\sqrt{D}$ (see Section~\ref{subsec: summary}). 
We measure the constant $c_{\epsilon}$ by tuning $c_{\epsilon}$ such that the curve of $\epsilon=c_{\epsilon}\cdot 2^{-B}/\sqrt{D}$ (the green curve) is higher than the curve of the 99.9\% quantile of the empirical errors. The result shows that when $c_{\epsilon}=5.75$, the curves match well. 

\section{related work}
\label{sec:related work}

\smallskip\noindent\textbf{Quantization.} 
A substantial body of literature on the quantization of high-dimensional vectors exists across various fields, including machine learning, computer vision, and data management~\cite{jegou2010product,additivePQ,ITQ,vaq,composite_quantization,ge2013optimized,lsq++, vaplusfile, splitvq, vafile,blink_scalar_quantization_graph,codesign2023,splitvq}.
We refer readers to comprehensive surveys and textbooks~\cite{learningtohash,surveyl2hash, ite_matsui_2018_pq_survey, dataseries_benchmark, revision1_bound, revision2_bound}. 
The existing studies about quantization can be divided into roughly two threads: (1) PQ and its variant~\cite{jegou2010product,additivePQ,ge2013optimized,lsq++, vaq} and (2) SQ and its variants~\cite{vafile, vaplusfile, blink_scalar_quantization_graph}.
We note that although the family of PQ and the family of SQ have highly diversified schemes, they can be presented in a unified framework: (1) in the index phase, they construct a quantization codebook and for each data vector, they find the nearest vector in it as the quantized data vector; and (2) in the query phase, they estimate the distance between a data vector and a query vector based on the quantized data vector.
Despite this, we observe that in existing literature~\cite{jegou2010product,douze2024faisslibrary,ge2013optimized,lsq,ite_matsui_2018_pq_survey}, PQ and SQ were seldom compared to each other~\footnote{The LVQ paper~\cite{blink_scalar_quantization_graph} compares PQ and SQ and finds that PQ does not have competitive performance with a moderate compression rate.}. Indeed, PQ and its variants are used mostly in the scenarios with large compression rates~\cite{jegou2010product,douze2024faisslibrary,ge2013optimized,lsq,ite_matsui_2018_pq_survey, learningtohash,surveyl2hash, composite_quantization} while SQ and its variants are used mostly in the scenarios with moderate compression rates~\cite{douze2024faisslibrary,PASE,milvus,blink_scalar_quantization_graph}.
According to our experimental results in Section~\ref{subsubsec: space-accuracy}, we find that with moderate compression rates (e.g., quantizing the vectors with $\ge 4$ bits per dimension), PQ does not outperform the classical SQ on many datasets.
In addition, as has been reported~\cite{blink_scalar_quantization_graph}, PQ is significantly slower than SQ in distance computation when the same number of bits are used.
On the other hand, with relatively large compression rates (e.g., quantizing the vectors with 1 or 2 bits per dimension), SQ can hardly produce reasonable accuracy. 
This may help explain why PQ and SQ were usually regarded as two separate lines of studies, i.e., each of them is capable of a certain scenario only.  
In contrast, in this study, we develop a method which excels in both scenarios. With the compression rates from 3x to 32x, our method stably outperforms all the methods empirically (Section~\ref{subsubsec: space-accuracy}). Moreover, it achieves the asymptotic optimality in theory (Section~\ref{subsec: summary}). 
In addition, it is worth noting that PQ and its variants are also used with a compression rate which is even larger than 32x. However, in this case, without re-ranking, they can only produce poor recall~\cite{jegou2010product,douze2024faisslibrary,ge2013optimized,lsq,ite_matsui_2018_pq_survey, learningtohash,surveyl2hash, composite_quantization} (e.g., <80\%), which deviates from the target of this study, i.e., achieving high recall without storing the raw vectors in RAM for re-ranking. 
For better comprehensiveness, we include the discussion on the way of using our method with a larger compression rate in Section~\ref{sec:conclusion and discussion} and leave more detailed study as future work.

\smallskip\noindent\textbf{ANN Query.}  ANN query is a key component of high-dimensional vector data management and has been widely supported by real-world systems~\cite{milvus,PASE,apple,manu,survey_vector_database_2024}.
Among the existing algorithms~\cite{malkov2018efficient,fu2019fast,li2019approximate,jegou2010product,ge2013optimized,datar2004locality,mtree,wang2024distancecomparisonoperatorsapproximate,yang2024effectivegeneraldistancecomputation}, the partition-based method IVF and the graph-based method HNSW have been deployed to the widest extent~\cite{milvus,PASE,survey_vector_database_2024,malkov2018efficient}.
For more methods, we refer readers to recent tutorials~\cite{tutorialThemis, tutorialXiao, tutorial_SIGMOD24}, surveys~\cite{metric_survey, survey_vector_database_2024, wang2023graph, aumullerrecent2023} and benchmarks~\cite{annbenchmark, graphbenchmark, li2019approximate, dobson2023scaling_billion_benchmark}.
Besides the ANN query, there have been growing interests in many advanced queries related to ANN~\cite{milvus,survey_vector_database_2024,apple}. 
For example, in real-world systems, besides vectors, a data object usually involves several attributes such as labels, numbers and strings. 
It is often the case that users target to find the data objects which satisfy some constraints on the attributes and have their vector nearest to the user's query vector. 
These questions are usually referred to as attribute-filtering ANN query~\cite{acorn, serf,milvus,apple,survey_vector_database_2024,PASE}.
Besides, due to the recent development in information retrieval, there has also been a trend in multi-vector search~\cite{colbert,colbertv2,survey_vector_database_2024}.
In particular, these studies map a document to multiple high-dimensional vectors. During querying, they also map a query document to multiple vectors and search for the most relevant documents via a new similarity metric called \emph{MaxSim}, which is aggregated from the inner products between query and data vectors~\cite{colbert,colbertv2}.
We note that in this paper, we present a method of vector compression and support to estimate inner product and squared Euclidean distances unbiasedly based on the compressed vectors.
Our method can be used in these tasks for vector compression and distances/inner-product estimation seamlessly. 

\smallskip\noindent\textbf{Random Projection.} 
Random projection is a fundamental technique that has wide applications~\cite{dasgupta2008random,datar2004locality,rabitq,mose_lsh,adsampling, tao2010efficient, sun2014srs, huang2015query}. 
The effectiveness of random projection is closely related to the seminal Johnson-Lindenstrauss (JL) Lemma~\cite{johnson1984extensions}.
In particular, JL Lemma states that projecting a vector onto the space of $O(\epsilon^{-2}\log (1/\delta))$ dimensions suffices to provide an error bound of $\epsilon$ with the probability of at least $1-\delta$~\footnote{The error bound includes a multiplicative error bound of Euclidean distances and an additive error bound of inner product.}. 
A theoretical study proves that JL Lemma is asymptotically optimal in terms of the trade-off between the number of \textit{dimensions} and the error bound~\cite{larsen2017optimality}. 
For more applications and theoretical conclusions, we refer readers to a comprehensive introduction~\cite{jlintro}.
Moreover, several studies further investigate on the trade-off between the number of \textit{bits} for representing a vector and the error bound of inner product/distance estimation~\cite{2017_focs_additive_error,SIAM_JOC_2022_indyk,bit_stoc_1998}. 
In particular, they prove that compressing a vector into a short code with $O(\epsilon^{-2}\log (1/\delta))$ bits suffices to guarantee an error bound of $\epsilon$. 
As a comparison, trivially applying JL Lemma and quantizing the real value of each dimension with SQ requires $O(\log \frac{1}{\epsilon})$ bits per dimension~\cite{2017_focs_additive_error}.
Furthermore, it is observed in~\cite{2017_focs_additive_error}, for the first time, that when the required accuracy is high ($\epsilon$ is small, $\epsilon^{-2} \log (1/\delta)>D$), the aforementioned result can be further improved (see Section~\ref{subsec: challenges}). 
However, as has been discussed in Section~\ref{subsec: summary}, the algorithmic proof in this study is hardly practically applicable.

\section{conclusion}
\label{sec:conclusion and discussion}

In conclusion, in this study, we propose a novel quantization algorithm which extends the RaBitQ method.
It supports to compress high-dimensional vectors with a moderate compression rate such that it produces promising recall without re-ranking. 
The method has consistent advantages in both empirical accuracy and theoretical guarantees over SQ and its variants. 
In addition, it supports efficient distance comparisons by first estimating a distance based on the first bits of the quantization codes, which helps prune many candidates from the distance computation based on the full codes. 
Extensive experiments verify its effectiveness in practice and the alignment between the empirical performance and the theoretical analysis. 

We would like to discuss on the following applications and extensions. 
(1) Like the original RaBitQ, the method can be adapted to support the estimation of inner product and cosine similarity (see \cite{rabitq}). 
(2) When targeting to compress vectors with a compression rate of >32x (i.e., we use fewer than 1 bit per dimension for quantizing a vector), we could first apply random projection for dimension reduction and apply the original RaBitQ method for binarization. 
The details are left in Appendix~\ref{rabitq with >32 compression rate} due to the page limit.
(3) Our method may bring benefits to other scenarios of ANN search (e.g., we may put the first bits of the quantization codes in RAM, leave the remaining bits in SSDs and conduct distance comparisons like what we do in Section~\ref{subsec: distance comparison with rabitq}).
(4) Quantization is also a fundamental component in machine learning systems. The practicality and optimality of RaBitQ implies its potential of optimizing neural network inferencing and training.

\bibliographystyle{ACM-Reference-Format}
\bibliography{main}

\appendix
\section*{appendix}

\section{The Proof of Lemma~\ref{lemma: quantization codes}}
\label{appendix: lemma 3.1}
\begin{proof}
    Note that 
    \begin{align}
        \| \mathbf{y}-t\mathbf{o}'\|^2 &=\| \mathbf{y}\|^2 + t^2 - 2t\left< \mathbf{o',y}\right>
        \\ \| \mathbf{\bar y}-t\mathbf{o}'\|^2 &=\| \mathbf{\bar y}\|^2 + t^2 - 2t\left< \mathbf{o',\bar y}\right>
    \end{align}
    Then to prove that $\exists t>0$ such that $\| \mathbf{y}-t\mathbf{o}'\|^2 \ge \| \mathbf{\bar y}-t\mathbf{o}'\|^2, \forall \mathbf{y}\in \mathcal{G}$, it suffices to prove that 
    \begin{align}
        \| \mathbf{y}\|^2 - \| \mathbf{\bar y}\|^2- 2t\left< \mathbf{o}',\mathbf{y}\right>+ 2t\left< \mathbf{o}', \mathbf{\bar y}\right> \ge 0
    \end{align}
    Let $t=\frac{\| \mathbf{\bar y}\|^2}{\left<\mathbf{o}',\mathbf{\bar y} \right>}$. We derive from the left hand side as follows.
    \begin{align}
        &\| \mathbf{y}\|^2 - \| \mathbf{\bar y}\|^2- 2t\left< \mathbf{o}',\mathbf{y}\right>+ 2t\left< \mathbf{o}',\mathbf{\bar y}\right>
        \\ = &\| \mathbf{y}\|^2 - \|\mathbf{\bar y}\|^2 -2\frac{\|\mathbf{\bar y}\|^2 }{\left< \mathbf{o}',\mathbf{\bar y}\right>}\cdot \left< \mathbf{o}',\mathbf{y}\right> +2\|\mathbf{\bar y}\|^2 \label{proof 3.1: plugin t}
        \\ = &\| \mathbf{y}\|^2 + \|\mathbf{\bar y}\|^2 -2\frac{\|\mathbf{\bar y}\|^2 }{\left< \mathbf{o}',\mathbf{\bar y}\right>}\cdot \left< \mathbf{o}',\mathbf{y}\right>  \label{proof 3.1: transform 1}
        \\ = &\| \mathbf{y}\|^2 + \|\mathbf{\bar y}\|^2 -2\frac{\|\mathbf{\bar y}\|\|\mathbf{y}\| }{\left< \mathbf{o}',\frac{\mathbf{\bar y}}{\|\mathbf{\bar y}\|}\right>}\cdot \left< \mathbf{o}',\mathbf{\frac{y}{\|\mathbf{y}\|}}\right>  \label{proof 3.1: transform 2}
        \\ \ge &\| \mathbf{y}\|^2 + \|\mathbf{\bar y}\|^2 -2\|\mathbf{\bar y}\|\|\mathbf{y}\| \label{proof 3.1: apply ineq}
        \\ = &\left( \| \mathbf{\bar y}\|-\|\mathbf{y}\|\right) ^2 \ge 0 \label{proof 3.1: finish}
    \end{align}
    where (\ref{proof 3.1: plugin t}) plugs in $t=\frac{\| \mathbf{\bar y}\|^2}{\left<\mathbf{o}',\mathbf{\bar y} \right>}$. (\ref{proof 3.1: transform 1}) and (\ref{proof 3.1: transform 2}) rewrite the form of (\ref{proof 3.1: plugin t}). (\ref{proof 3.1: apply ineq}) applies the fact that $\mathbf{\bar y} =\argmax_{\mathbf{y} \in \mathcal{G}} \left<\frac{\mathbf{ y}}{\|\mathbf{y}\|}, \mathbf{o}' \right>$, i.e., $\forall \mathbf{y} \in \mathcal{G}, \left<\frac{\mathbf{\bar y}}{\|\mathbf{\bar y}\|}, \mathbf{o}' \right> \ge \left<\frac{\mathbf{ y}}{\|\mathbf{y}\|}, \mathbf{o}' \right>$ and $\left<\frac{\mathbf{\bar y}}{\|\mathbf{\bar y}\|}, \mathbf{o}' \right> > 0$. (\ref{proof 3.1: finish}) finishes the proof.
\end{proof}

\section{The Proof of Theorem~\ref{theorem: main}}
\label{appendix: proof main theorem}
To make the paper self-contained, we restate the tail bound of the coordinates of the random vector which follows uniform distribution on the unit sphere $\mathbb{S}^{D-1}$ in the $D$-dimensional space.

\begin{lemma}(\cite{vershynin_2018})
\label{lemma: distribution}
    For a random vector $\mathbf{x}=(\mathbf{x}[1], \mathbf{x}[2],...,\mathbf{x}[D]) $ which follows the uniform distribution on the unit sphere $\mathbb{S}^{D-1}$ in the $D$-dimensional space, the tail bound of its coordinates is given as 
    \begin{align}
        \mathbb{P}\left\{ \left| \mathbf{x}[i]\right| > \frac{t}{\sqrt {D} }     \right\}  \le 2\exp \left( -c_0 t^2 \right) 
    \end{align}
    where $c_0$ is a constant, $i=1,2,...,D$.
\end{lemma}

We next prove Theorem~\ref{theorem: main} by first proving 
Lemma~\ref{lemma: B.1} and Lemma~\ref{lemma B.3}. 

\begin{lemma}
    \label{lemma: B.1}
    Let $\mathbf{o}$ be a unit vector and $\mathbf{\bar o}$ be its quantized vector in $\mathcal{G}_r$, $\mathbf{o}'=P^{-1}\mathbf{o}$. Then
    \begin{align}
        \sqrt{1-\left< \mathbf{o},\mathbf{\bar o}\right>^2}=\min_{t>0, \mathbf{y}\in \mathcal{G}} \| \mathbf{o}'-t\mathbf{y}\|
    \end{align}    
\end{lemma}
\begin{proof}
    We first prove that $\forall t>0,\mathbf{y}\in \mathcal{G}, \| \mathbf{o}'-t\mathbf{y}\|^2 \ge 1-\left< \mathbf{o},\mathbf{\bar o}\right>^2$ as follows.
    \begin{align}
        \| \mathbf{o}'-t\mathbf{y}\|^2&=\| \mathbf{o}'\|^2-2 t\left<\mathbf{o}', \mathbf{y}\right> + t^2 \| \mathbf{y}\|^2
        \\ &= 1 - \left< \mathbf{o},\mathbf{\bar o}\right>^2 + \left< \mathbf{o},\mathbf{\bar o}\right>^2 -2 t\left<\mathbf{o}', \mathbf{y}\right> + t^2 \| \mathbf{y}\|^2 \label{eq: appendix B o_prime}
        \\ &= 1 - \left< \mathbf{o},\mathbf{\bar o}\right>^2 + \left< \mathbf{o}',\frac{\mathbf{\bar y}}{\|\mathbf{\bar y}\|}\right>^2 -2 t\left<\mathbf{o}', \mathbf{y}\right> + t^2 \| \mathbf{y}\|^2 \label{eq: appendix B definition of o_prime}
        \\ &\ge 1 - \left< \mathbf{o},\mathbf{\bar o}\right>^2 + \left< \mathbf{o}',\frac{\mathbf{y}}{\|\mathbf{y}\|}\right>^2 -2 t\left<\mathbf{o}', \mathbf{y}\right> + t^2 \| \mathbf{y}\|^2 \label{eq: appendix B relax y}
        \\ & =1 - \left< \mathbf{o},\mathbf{\bar o}\right>^2 + \left( \left< \mathbf{o}',\frac{\mathbf{y}}{\|\mathbf{y}\|}\right>-t\|\mathbf{y}\|\right)^2 \label{eq: appendix B finish 1}
        \\ & \ge 1 - \left< \mathbf{o},\mathbf{\bar o}\right>^2 \label{eq: appendix B finish 2}
    \end{align}
    where (\ref{eq: appendix B o_prime}) is because $\mathbf{o}'$ is a unit vector. (\ref{eq: appendix B definition of o_prime}) applies $P^{-1}$ to both sides of the third term in (\ref{eq: appendix B o_prime}). (\ref{eq: appendix B relax y}) is because $\mathbf{\bar y}=\argmax_{\mathbf{y} \in \mathcal{G}} \left<\mathbf{y}/\|\mathbf{y}\|, \mathbf{o}' \right>$ and $\mathcal{G}$ is symmetric, i.e., $\forall \mathbf{y}\in \mathcal{G}$, we have $\mathbf{-y} \in \mathcal{G}$. 
    Then we note that when $\mathbf{y}=\mathbf{\bar y}$ and $t=\frac{1}{\|\mathbf{\bar y}\|}\left< \mathbf{o}',\frac{\mathbf{\bar y}}{\|\mathbf{\bar y}\|}\right>$, the equality holds. Taking the square root of both sides finishes the proof.
\end{proof}

\begin{lemma}
    \label{lemma B.3}
    For any $L > 0$, we have
    \begin{align}
        \mathbb{P} \left\{ \sqrt{1-\left< \mathbf{o},\mathbf{\bar o}\right>^2} > \frac{L}{2^{B}} + \frac{c_1}{\sqrt{\delta}}\cdot \exp \left( - \frac{c_0}{2} L^2\right)  \right\} < \delta \label{eq: conclusion lemma B.2}
    \end{align}
    where $c_0$ and $c_1$ are absolute constants.
\end{lemma}

\begin{proof}
    Let $f(x)= \mathrm{sgn} (x) \cdot \min \left( |x|, \frac{L}{\sqrt{D}}\right) $ where $\mathrm{sgn}(x) = -1$ if $x<0$ and $\mathrm{sgn}(x) = +1$ if $x\ge 0$. We let $f(\mathbf{o}')$ denote the vector whose $i$th coordinate is $f(\mathbf{o}'[i])$. Next we start the derivation from $\sqrt{1-\left< \mathbf{o},\mathbf{\bar o}\right>^2}$ as follows.
    \begin{align}
        \sqrt{1-\left< \mathbf{o},\mathbf{\bar o}\right>^2} &= \min_{t>0, \mathbf{y}\in \mathcal{G}} \| \mathbf{o}'-t\mathbf{y}\| \label{eq: lemma B.2 apply lemma B.1}
        \\ &\le \min_{\mathbf{y}\in \mathcal{G}} \left\| \mathbf{o}'-t_0\mathbf{y} \right\| \label{eq: lemma B.2 relax t_0}
        \\ &= \min_{\mathbf{y}\in \mathcal{G}} \left\| \mathbf{o}'-f(\mathbf{o}') +f(\mathbf{o}') -t_0\mathbf{y} \right\| \label{eq: lemma B.2 triangle 1}
        \\ &\le \min_{\mathbf{y}\in \mathcal{G}} \left\| \mathbf{o}'-f(\mathbf{o}') \right\|+\left\|f(\mathbf{o}') -t_0\mathbf{y} \right\| \label{eq: lemma B.2 triangle 2}
        \\ &= \left\| \mathbf{o}'-f(\mathbf{o}') \right\| + \min_{\mathbf{y}\in \mathcal{G}} \left\|f(\mathbf{o}') -t_0\mathbf{y} \right\| \label{eq: lemma B.2 separate}
    \end{align}
    Here $t_0:=\frac{L}{2^{B-1}\sqrt{D}}$. (\ref{eq: lemma B.2 apply lemma B.1}) applies Lemma~\ref{lemma: B.1}. (\ref{eq: lemma B.2 relax t_0}) relaxes the minimum over $t>0$ to a specific $t_0$. 
    (\ref{eq: lemma B.2 triangle 1}) and (\ref{eq: lemma B.2 triangle 2}) apply triangle's inequality.
    (\ref{eq: lemma B.2 separate}) holds because $\left\| \mathbf{o}'-f(\mathbf{o}') \right\|$ is independent of $\mathbf{y}$.
    We next analyze these two terms in (\ref{eq: lemma B.2 separate}) separately. Let us first analyze the expected value of $\left\| \mathbf{o}'-f(\mathbf{o}') \right\|^2$ as follows. 
    \begin{align}
        \mathbb{E} \left[ \left\| \mathbf{o}'-f(\mathbf{o}') \right\|^2\right]
        =&\mathbb{E}\left[ \sum_{i=1}^{D}  |\mathbf{o}'[i]-f(\mathbf{o}'[i])|^2\right]
        \\=&D\cdot \mathbb{E}\left[ |\mathbf{o}'[1]-f(\mathbf{o}'[1])|^2\right] \label{eq: lemma B.2 linear expectation}
        \\=& D\cdot \int_{0}^{+\infty} \mathbb{P} \left\{ |\mathbf{o}'[1]-f(\mathbf{o}'[1])|^2>t \right\}\mathrm{d}t \label{eq: lemma B.2 elementary}
    \end{align}
    \begin{align}
        =& D\cdot \int_{0}^{+\infty} \mathbb{P} \left\{ |\mathbf{o}'[1]|>\sqrt{t} + \frac{L}{\sqrt{D}} \right\}\mathrm{d}t \label{eq: lemma B.2 verify by f(x)}
        \\< & D\cdot \int_{0}^{+\infty}2\exp \left( -c_0 (\sqrt{Dt} + L)^2\right)\mathrm{d}t  \label{eq: lemma B.2 plugin tail bound}
        \\\le &D\cdot e^{-c_0L^2} \int_{0}^{+\infty}2\exp \left( -c_0Dt\right)\mathrm{d}t \label{eq: lemma B.2 simplify 1}
        \\= &D\cdot e^{-c_0L^2} \frac{2}{c_0D}=\frac{2}{c_0}\cdot e^{-c_0L^2} \label{eq: lemma B.2 simplify 2}
    \end{align}
    (\ref{eq: lemma B.2 linear expectation}) is because the linearity of expectation and the fact that $\mathbf{o}'[i]$'s are identically distributed to each other. (\ref{eq: lemma B.2 elementary}) is an elementary property of non-negative random variables. (\ref{eq: lemma B.2 verify by f(x)}) can be verified by the definition of $f(x)$. (\ref{eq: lemma B.2 plugin tail bound}) applies Lemma~\ref{lemma: distribution}.
    (\ref{eq: lemma B.2 simplify 1}) holds because $(\sqrt{Dt}+L)^2 > Dt + L^2$. (\ref{eq: lemma B.2 simplify 2}) is by elementary calculus.
    Next by applying Markov's inequality~\cite{vershynin_2018}, we derive the tail bound of $\left\| \mathbf{o}'-f(\mathbf{o}') \right\|^2$ as follows.
    \begin{align}
         &\mathbb{P}\left\{ \left\| \mathbf{o}'-f(\mathbf{o}') \right\|^2 \ge \frac{2}{c_0\delta} \cdot \exp \left( -c_0L^2\right) \right\}
        \\ \le & \frac{c_0\delta}{2} \cdot \exp \left( c_0L^2\right) \cdot \mathbb{E} \left[ \left\| \mathbf{o}'-f(\mathbf{o}') \right\|^2\right]
        \\ < & \frac{c_0\delta}{2} \cdot \exp \left( c_0L^2\right) \cdot \frac{2}{c_0}\cdot e^{-c_0L^2}= \delta \label{46}
    \end{align}
    Then we analyze the upper bound of the second term in (\ref{eq: lemma B.2 separate}). 
    Note that based on the definition of $f(x)$, every coordinate of $f(\mathbf{o}')$ is within the range of $[-\frac{L}{\sqrt{D}}, +\frac{L}{\sqrt{D}}]$. 
    Let us choose $\mathbf{y}$ from the set of grids $\mathcal{G}$ such that every coordinate $t_0\mathbf{y}[i]$ is closest to $f(\mathbf{o}'[i])$.
    Recall that $t_0=\frac{L}{\sqrt{D}}\cdot \frac{1}{2^{B-1}}$ and $\mathbf{y}[i]$ ranges from $(-2^{B-1} + \frac{1}{2})$ to $(+2^{B-1} - \frac{1}{2})$. 
    Then for every dimension $i$, we have $|f(\mathbf{o}'[i])-t_0\mathbf{y}[i]| \le \frac{1}{2^{B}}\frac{L}{\sqrt{D}}$.
    Thus, we have 
    \begin{align}
        \min_{\mathbf{y}\in \mathcal{G}} \left\|f(\mathbf{o}') -t_0\mathbf{y} \right\| \le  \frac{L}{2^B} \label{47}
    \end{align}
    Based on the analysis above, we prove (\ref{eq: conclusion lemma B.2}) as follows. 
    \begin{align}
        & \mathbb{P} \left\{ \sqrt{1-\left< \mathbf{o},\mathbf{\bar o}\right>^2} > \frac{L}{2^{B}} + \sqrt{\frac{2e^{-c_0 L^2}}{c_0\delta}}\right\}
        \\ \le & \mathbb{P} \left\{ \left\| \mathbf{o}'-f(\mathbf{o}') \right\| + \min_{\mathbf{y}\in \mathcal{G}} \left\|f(\mathbf{o}') -t_0\mathbf{y} \right\| > \frac{L}{2^{B}} + \sqrt{\frac{2e^{-c_0 L^2}}{c_0\delta}}\right\} \label{eq: lemma B.2 49 apply}
        \\ \le & \mathbb{P} \left\{ \left\| \mathbf{o}'-f(\mathbf{o}') \right\| + \frac{L}{2^{B}} > \frac{L}{2^{B}} + \sqrt{\frac{2e^{-c_0 L^2}}{c_0\delta}}\right\} \label{eq: lemma B.2 50 apply}
        \\ = & \mathbb{P} \left\{ \left\| \mathbf{o}'-f(\mathbf{o}') \right\| > \sqrt{\frac{2e^{-c_0 L^2}}{c_0\delta}}\right\} \le \delta  \label{51}
    \end{align}
    where (\ref{eq: lemma B.2 49 apply}) applies (\ref{eq: lemma B.2 separate}). (\ref{eq: lemma B.2 50 apply}) applies (\ref{47}). (\ref{51}) applies (\ref{46}).
\end{proof}

Based on the lemmas above, we prove Theorem~\ref{theorem: main} as follows. 
\begin{proof}
    For simplifying the notations, let us define the following events. 
    \begin{align}
        E_A: & \left| \frac{\left< \mathbf{\bar o}, \mathbf{q}  \right> }{\left< \mathbf{\bar o},\mathbf{o}  \right> } -\left< \mathbf{o,q} \right>   \right| \le \sqrt{\frac{{1 - \left< \mathbf{\bar o}, \mathbf{o}  \right>^2}}{\left< \mathbf{\bar o}, \mathbf{o}  \right>^2 }}\cdot \sqrt{\frac{\log(1/\delta)}{c_0 (D-1)}} 
        \\ E_B: & \sqrt{\frac{{1 - \left< \mathbf{\bar o}, \mathbf{o}  \right>^2}}{\left< \mathbf{\bar o}, \mathbf{o}  \right>^2 }} \le \frac{L}{2^{B}} + \frac{c_1}{\sqrt{\delta}}\cdot \exp \left( - \frac{c_0}{2} L^2\right)
    \end{align}
    In particular, to provide an upper bound for the error of the estimator, we target to prove that 
    \begin{align}
        \mathbb{P} \left\{ E_A\ \mathrm{and}\ E_B\right\} \ge 1-2\delta
    \end{align}
    Based on the union bound, it suffices to prove that 
    \begin{align}
        \mathbb{P} \left\{ \lnot E_A\right\} \le \delta \ \mathrm{and} \ \mathbb{P} \left\{ \lnot E_B\right\} \le \delta
    \end{align}
    Note that $\mathbb{P} \left\{ \lnot E_A\right\} \le \delta$ holds due to Lemma~\ref{lemma: restate} and $\mathbb{P} \left\{ \lnot E_B\right\} \le \delta$ holds because of Lemma~\ref{lemma B.3} (Lemma~\ref{lemma B.3} provides an upper bound for $\sqrt{1-\left< \mathbf{o},\mathbf{\bar o}\right>^2}$ and implies that $\left< \mathbf{o},\mathbf{\bar o}\right>$ is lower bounded by $1/2$). Thus, with the probability of at least $1-2\delta$, we have 
    \begin{align}
        \left| \frac{\left< \mathbf{\bar o}, \mathbf{q}  \right> }{\left< \mathbf{\bar o},\mathbf{o}  \right> } -\left< \mathbf{o,q} \right>   \right| \le \left[ \frac{L}{2^{B}} + \frac{c_1}{\sqrt{\delta}}\cdot \exp \left( - \frac{c_0}{2} L^2\right)\right]\cdot \sqrt{\frac{\log(1/\delta)}{c_0 (D-1)} }
    \end{align}
    To bound the right hand side by $\epsilon$, it suffices to let
    \begin{align}
        L= & \Theta \left( \log \left[ \frac{1}{D} \cdot \frac{(1/\delta)\cdot \log (1/\delta)}{\epsilon^2 }\right]\right)
        \\ B= & \Theta \left( \log \left[ \frac{1}{D} \cdot \frac{\log (1/\delta)}{\epsilon^2 }\right] + \log L\right) 
        \\ = &\Theta \left( \log \left[ \frac{1}{D} \cdot \frac{\log (1/\delta)}{\epsilon^2 }\right]\right) 
    \end{align} 
\end{proof}

\section{Using RaBitQ with >32x Compression Rate}
\label{rabitq with >32 compression rate}
Let $P_d$ be the random orthogonal matrix which projects a $D$-dimensional vector to $d$-dimensional space.   
We present the details of the estimation for $\| \mathbf{o}_r-\mathbf{q}_r\|^2$ as follows.
\begin{align}
&\| \mathbf{o}_r-\mathbf{q}_r \|^2 \\ 
    = &\| {\mathbf{o}_r -\mathbf{c} }\|^2 + \| \mathbf{q}_r- \mathbf{c} \|^2 -2 \cdot \left< \mathbf{o}_r-\mathbf{c}, \mathbf{q}_r-\mathbf{c} \right>  \\
    \approx & \| {\mathbf{o}_r -\mathbf{c} }\|^2 + \| \mathbf{q}_r- \mathbf{c} \|^2 -\frac{2D}{d} \cdot 
    \left< P_d(\mathbf{o}_r-\mathbf{c}), P_d(\mathbf{q}_r-\mathbf{c}) \right>  \label{larger compression 17}\\
    = & \| {\mathbf{o}_r -\mathbf{c} }\|^2 + \| \mathbf{q}_r- \mathbf{c} \|^2 -\frac{2D}{d} \cdot \|P_d(\mathbf{o}_r-\mathbf{c})\|\cdot \|P_d(\mathbf{q}_r-\mathbf{c})\| \nonumber \\ 
    &\cdot 
    \left< \frac{P_d(\mathbf{o}_r-\mathbf{c})}{\|P_d(\mathbf{o}_r-\mathbf{c})\|}, \frac{P_d(\mathbf{q}_r-\mathbf{c})}{\|P_d(\mathbf{q}_r-\mathbf{c})\|} \right> \\ \approx  &\| {\mathbf{o}_r -\mathbf{c} }\|^2 + \| \mathbf{q}_r- \mathbf{c} \|^2 -2 \cdot \|\mathbf{o}_r-\mathbf{c}\|\cdot \|\mathbf{q}_r-\mathbf{c}\| \nonumber \\ 
    &\cdot 
    \left< \frac{P_d(\mathbf{o}_r-\mathbf{c})}{\|P_d(\mathbf{o}_r-\mathbf{c})\|}, \frac{P_d(\mathbf{q}_r-\mathbf{c})}{\|P_d(\mathbf{q}_r-\mathbf{c})\|} \right> \label{larger compression 19}
\end{align}
Here (\ref{larger compression 17}) and (\ref{larger compression 19}) are based on Johnson-Lindenstrauss Lemma~\cite{johnson1984extensions}.
$\left< \frac{P_d(\mathbf{o}_r-\mathbf{c})}{\|P_d(\mathbf{o}_r-\mathbf{c})\|}, \frac{P_d(\mathbf{q}_r-\mathbf{c})}{\|P_d(\mathbf{q}_r-\mathbf{c})\|} \right>$ is the inner product between two unit vectors in $d$-dimensional space, whose estimation can be realized with RaBitQ.

\end{document}